\documentclass[11pt]{extarticle}
\usepackage{geometry}                
\usepackage{graphicx}
\usepackage{amssymb}
\usepackage{epstopdf}
\usepackage{dsfont}
\usepackage{mathrsfs}
\usepackage{scrextend}
\usepackage{bm}
\usepackage{mathtools}
\usepackage{lscape}
\newtheorem{theorem}{Theorem}
\newtheorem{lemma}[theorem]{Lemma}
\newtheorem{proposition}[theorem]{Proposition}
\newtheorem{corollary}[theorem]{Corollary}

\begin{document}

\newcommand\eqquery{\stackrel{\mathclap{\normalfont\mbox{?}}}{=}}

\parskip=5pt     

\newcommand{\riota}{\mathrm{\rotatebox[origin=c]{180}{$\iota$}}}

\begin{center}
{\Large\bf Is a particle an irreducible representation of the Poincar\'e group?}

Adam Caulton (adam.caulton@philosophy.ox.ac.uk), 29 September 2024
\end{center}

\begin{abstract}
The claim that a particle is an irreducible representation of the Poincar\'e group---what I call \emph{Wigner's identification}---is now, decades on from Wigner's (1939) original paper, so much a part of particle physics folklore that it is often taken as, or claimed to be,  a definition. My aims in this paper are to: (i) clarify, and partially defend, the guiding ideas behind this identification; (ii) raise objections to its being an adequate definition; and (iii) offer a rival characterisation of particles. My main objections to Wigner's identification appeal to the problem of interacting particles, and to alternative spacetimes. I argue that the link implied in Wigner's identification, between a spacetime's symmetries and the generator of a particle's space of states, is  at best misleading, and that there is no good reason to link  the generator of a particle's space of states to symmetries of any kind. I propose an alternative characterisation of particles, which captures both the relativistic and non-relativistic setting. I further defend this proposal by appeal to a theorem which links the decomposition of Poincar\'e generators into purely orbital and spin components with canonical algebraic relations between position, momentum and spin.
\end{abstract}

\tableofcontents

\section{Wigner's identification} \label{Section1}

Ask any particle physicist, `What is a particle?' and you are likely to receive the reply, `a particle is a representation of the Poincar\'e group', or perhaps, `an elementary particle is an irreducible representation of the Poincar\'e group'. To quote a reliable textbook, 
\begin{quote}
`Ever since the fundamental paper of Wigner on the irreducible representations of the Poincar\'e group, it has been a (perhaps implicit) definition in physics that an elementary particle ``is'' an irreducible representation of the group, $G$, of ``symmetries of nature''.'
 Ne'eman and Sternberg (1991, p.~327).
\end{quote}
The group $G$ that Ne'eman and Sternberg have in mind here is the direct product of the Poincar\'e group, the group of `external'  symmetries of Minkowski spacetime, with the group of `internal' symmetries associated with the Standard Model of particle physics, namely $SU(3)\times SU(2)\times U(1)$. This highly abstract answer to the seemingly  simple question, `What is a particle?' I will call \emph{Wigner's identification}, after Wigner's (1939) celebrated treatment. But my main focus here is on the `external symmetries', namely the Poincar\'e group. I don't wish to take issue with the claim that these `internal' symmetries are also part of a full characterisation of particles, but I do wish to take issue with the focus on the Poincar\'e group in, as it were, the `external' part of the characterisation.

My aim in this paper is to explain the idea behind Wigner's identification, investigate the commonly deployed argument for it, criticise it, and finally defend an alternative, which is largely inspired by Foldy (1956). What is by now a cluster of similar arguments in the physics literature for Wigner's identification traces back also to Wigner (1939), and makes essential use of what is usually known as Wigner's theorem: roughly, the result in quantum mechanics that transition-probability-preserving transformations are represented by unitary or anti-unitary operators. I expound this argument in this section, section \ref{Section1}, after some discussion of what Wigner's identification really means, and how it is usually deployed. Then, in section \ref{Section2}, I offer reasons to doubt that Wigner's identification can be correct. For there are situations in which particle-talk seems legitimate, or worth vindicating, but for which Wigner's identification would then fail: these situations are essentially ones in which particles interact, or in which the background spacetime is not highly symmetric (as in Minkowski spacetime). In section \ref{Section3} I defend a rival characterisation of particles, which is  most familiar from nonrelativistic quantum mechanics, but  (so I shall argue) applies equally well to the relativistic regime.

A note about jargon. I call it Wigner's \emph{identification} because this term seems to me most neutral on whether or not it was or is intended as a definition. As I will argue, it could not be a definition because it is not even extensionally adequate, since it fails to capture much of what we should think of as particles. I describe the search for an alternative as a search for a rival \emph{characterisation} because this term seems to me closer to definition, without committing to anything like the characterisation being analytic, or applicable to any conceptually possible situation. What is wanted, and what I aim to provide, is a characterisation which captures all and only what we should think of as particles in relativistic and nonrelativistic regimes, and in classical or quantum regimes.

Particles are not fundamental. This is by now the received view, and I do not disagree with it. Indeed, an important aspect to my proposed rival working relies on particle-talk being at best approximate, and nothing more than a useful conceptual carving of the quantum field in certain regimes. However, I do question some of the more extreme eliminativism about particles---especially in the context of interacting field theories---that have relatively recently been expressed (Earman \& Fraser 2006; Fraser 2008). In fact a large plank of my case against Wigner's identification depends on accepting particle talk as legitimate  in the presence of interactions; more on this in section \ref{Section2}.

For the remainder of this section, I will articulate what I think is behind the idea of Wigner's identification (\ref{WignerMeans}), briefly outline the contexts in which it is used (\ref{WignerUses}), and expound and partially reconstruct Wigner's argument (\ref{WignerArgues}).

\subsection{What does Wigner's identification mean?} \label{WignerMeans}

What could it even mean to say that a particle is an irreducible representation of the Poincar\'e group? The claim seems to commit some kind of basic category error. How could a particle, a constituent of the physical world, be something as abstract and mathematical as an irreducible representation of any group? Many are familiar with the mathematician's `if'---a word that means `if and only if' in the mouths of many mathematicians giving a definition. Here we have an instance of the physicist's `is'---a word that means `is represented by' in the mouths of physicists suppressing, for convenience, the distinction between the physical world and the mathematics that represents it. (Other examples include claims such as `the state of the system is $\psi$' or `spacetime is a smooth manifold'.)

So a particle is \emph{represented by} an irreducible representation of the Poincar\'e group. How? At this point we must take measures to deal with the two notions of `representation' in play. One notion is purely mathematical, the subject of group representation theory. In this subsection I will call representations in this sense \emph{reps}, and irreducible representations \emph{irreps}. The second notion is a relation, concerning the way that mathematical structures can describe, denote or otherwise stand for physical systems; I will continue to use the full word `representation' for this relation.

Any rep is a map $U:G\to\mathfrak{B}(\mathcal{H})$ from some group $G$ into the bounded linear operators on some Hilbert space, the carrier space for the representation. The map should send elements of the abstract group to operators such that the group structure of $G$ is preserved: that is, $U(g\circ g') = U(g)U(g')$ for all $g,g'\in G$. $U$ is a unitary representation iff each $U(g)$ is a unitary operator. But our central concern here is \emph{projective} unitary representations, and this permits a weakening of the defining condition to $U(g\circ g') = \omega(g, g')U(g)U(g')$, where $\omega(g, g')$ is a complex number, determined by $g$ and $g'$, on the unit circle, so that $|\omega(g, g')| = 1$. A rep is \emph{irreducible}, i.e.~an irrep, iff the only subspaces of the carrier space $\mathcal{H}$ which are invariant under all of the $U(g)$ are the trivial ones, i.e.~the zero subspace and $\mathcal{H}$ itself. Thus a rep is an irrep iff $G$ exhausts or generates, via $U$, the entire carrier space $\mathcal{H}$.

That is a broad outline of the mathematics. But what exactly is being represented? The carrier space $\mathcal{H}$ is the (mathematical representative of the) space of states of the system in question. Thus an irrep represents a system by representing that system's space of possibilities. In the case of particles, Wigner's identification is saying that: (i) a particle is represented by its space of possible states; and that (ii) this space of possible states is generated, via some irrep, by the transformations in the Poincar\'e group.

The general idea behind (i) is that any \emph{kind} of system may be characterised by the space of possible states for systems of that kind. This is a very powerful idea, and by no means needs to be limited to particles. For example, we might as well say a spinor is an irreducible representation of $SU(2)$, or that a scalar quantum field is an irreducible representation of the group generated by some appropriate Weyl algebra. And note that what Wigner's identification is doing is characterising a \emph{kind} of thing---not any specific instance of that kind. This is a very different idea from, say, specifying an object by offering a uniquely referring definite description.

But in fact an irrep can do more than generate the space of possible states for a system. If the relevant group is a Lie group, then we can and should further demand that any irrep be continuous, in the sense that $\lim_{g\to g'}U(g) = U(g')$ for all $g,g'\in G$. What is remarkable about this is that $U$ then induces a derivative map on $G$'s associated Lie algebra $\mathfrak{g}$---very roughly, infinitesimal transformations in the neighbourhood of the identity in $G$.\footnote{Less roughly, the Lie algebra $\mathfrak{g}$ of $G$ is the tangent space $T_eG$ at the identity $e$, equipped with a bracket $[\cdot, \cdot]: \mathfrak{g}\otimes \mathfrak{g}\to \mathfrak{g}$, whose properties are induced by the group composition function on $G$.} Each Lie algebra element generates a 1-parameter family of transformations in $G$ in exactly the same sense in which a tangent vector at some point $p$ on a smooth manifold generates a curve on that manifold which passes through $p$. If the irrep $U$ is continuous, then the full  1-parameter family of transformations in $G$ is sent to a 1-parameter family of unitaries on the carrier space $\mathcal{H}$. Now by Stone's Theorem, since this family of unitaries is continuous, it is generated by some self-adjoint operator. This operator is precisely the image of the  derivative map induced by $U$ on the corresponding Lie algebra element.

Thus a unitary irrep of a Lie group does not just produce a space of possible states for a system; it also produces a (Lie) algebra of self-adjoint operators, which may seen as generating the transformations on the carrier space. In quantum mechanics, self-adjoint operators lead a double life. On the one hand they are the generators of 1-parameter families of unitaries, as just described. On the other hand they are  \emph{quantities}, i.e.~determinable properties of the system in question. Thus, for example, momentum is both the generator of spatial translations and a quantity in its own right. In that case, a unitary irrep of a Lie group gives us an \emph{entire quantum theory:} for it gives us both a space of states and an algebra of quantities which acts irreducibly on that space. (In particular, if the Lie group in question includes translations in time, then the corresponding algebra contains a Hamiltonian whose form will be determined by its commutation relations between other elements in the algebra. Thus we have, among other things, a dynamics for our system.)

One attractive feature of this way of coming to a quantum theory is that it is  divorced from the idea of quantization---in which a quantum theory is obtained from some associated classical theory by a (typically non-algorithmic!)~procedure. A conceptual difficulty arising from thinking of quantization as the source of a quantum theory is that it seems to put the  quantum theory and its classical counterpart in the wrong order. Quantum mechanics doesn't come from classical mechanics; classical mechanics, insofar as it applies to the real world, emerges from quantum mechanics in some regime of approximation! On the conception just outlined above, the quantum theory comes directly from the choice of some Lie group and an investigation  of the irreps of that group on Hilbert spaces. Thus the quantum theory stands on its own two feet. 

Relatedly, it must also be pointed out that the game of choosing a Lie group and seeking irreducible representations can be played in a classical Hamiltonian setting too. In this setting, the concrete representatives of the group elements are not unitaries, linear transformations which preserve the inner product, but symplectomorphisms, a.k.a.~canonical transformations, which are diffeomorphisms on the classical phase space that preserve the symplectic form. The relevant map between the abstract group $G$ and symplectomorphisms is sometimes also called a representation, but called by Sudarshan and Mukunda (1974, ch.~14) a \emph{realization.} In the case of Lie groups, realizations can be constrained to be continuous, and this induces a map from the abstract group's Lie algebra to an algebra of classical quantities (that is, real-valued functions on the phase space) governed by the Poisson bracket. (Thus the abstract Lie algebra bracket is represented in the quantum setting by the commutator---divided by $i\hbar$---and in the classical setting by the Poisson bracket.) 
The weakening to projective unitary representations, justified by the invariance of the physical quantum state under a change of global phase, also has a counterpart for classical realizations. So the general idea behind Wigner's identification---the characterisation of a physical system in terms of its possible states, as generated, \emph{via} a representation or realization of some abstract Lie group---provides an origin for classical Hamiltonian theories just as much as for quantum theories.

This remarkable and powerful general idea is not something I wish to challenge. My specific concern is whether Wigner's identification chooses the \emph{right} Lie group to characterise particles. So let us turn to the Poincar\'e group. I will just briefly report well known results about the representation theory of the Poincar\'e group, stemming from Wigner's (1939) original treatment. To keep things brief, I will completely gloss over the wonderful sub-discipline of group representation theory known as `Mackey theory', tracing back to Mackey (1951) and having as one of its applications the irreps of the Poincar\'e group.

The Poincar\'e group may also be called the group of inhomogeneous restricted Lorentz transformations. The Lorentz transformations, comprising the group  $SO(3,1)$, are the transformations which preserve the Minkowski metric in a given vector space: these include spatial rotations on 3-space and Lorentz boosts (`pseudorotations' on spacetime). This group has 6 dimensions: 3 for spatial rotations and 3 for the Lorentz boosts. This group is formed of four disconnected components, since among the Lorentz transformations are parity transformations and time-reversal transformations (and transformations which invert both handedness and time-order), and such transformations are not continuously connected to the identity. The \emph{restricted} Lorentz group is the component connected to the identity, which preserve both handedness  (\emph{proper}) and time-order (\emph{orthochronous}). This is the group of transformations between relativistic inertial frames which preserves the spacetime origin of each frame.

We obtain the \emph{inhomogeneous} restricted Lorentz transformations by adding the 4 dimensions of translations in space and time. A generic restricted Lorentz transformation may be parametrised by an anti-symmetric $4\times4$ matrix $\omega_{\kappa\lambda} = -\omega_{\lambda\kappa}$, since this well encapsulates the 6 parameters determining the specific Lorentz transformation in question, which we may denote by $\Lambda(\omega)$. Since Lorentz transformations are linear, $\Lambda(\omega)$ is a matrix: or more exactly, a rank-(1, 1) tensor.  Further, let us denote a generic spacetime translation by the four-vector $a^\mu$, along which the translation is defined. Then a generic Poincar\'e transformation may be denoted by $(\Lambda(\omega), a)$. Its action on  four-position $x^\mu$ is
\begin{equation}
(\Lambda(\omega), a)(x^\mu) = \Lambda(\omega)^\mu_\nu x^\nu + a^\mu\ .
\end{equation}
Consideration of repeated Poincar\'e transformations reveals the composition law
\begin{equation}
(\Lambda(\omega'), a') \circ (\Lambda(\omega), a) = (\Lambda(\omega')\Lambda(\omega), \Lambda(\omega')(a) + a')\ .
\end{equation}
This gives the Poincar\'e group the form of a semidirect product of the group of spacetime translations and the restricted Lorentz group.

Since the Poincar\'e group is connected, it is entirely generated by exponentiating its associated Lie algebra. In terms of its concrete representation as unitary operators on some Hilbert space, translations by $a^\mu$ are given by $\exp(-\frac{i}{\hbar} a^\mu P_\mu)$, where $P_\mu$, the four momentum, is the generator of spacetime translations, and the Lorentz transformation $\Lambda(\omega)$ is given by $\exp(-\frac{i}{2\hbar}\omega_{\mu\nu}M^{\mu\nu})$, where $M^{\mu\nu}$ the relativistic angular momentum, is the generator of Lorentz transformations.

It is sometimes enlightening to express these generators with an explicit space/time split, and I will do that here. So defining $H := P_0$, $J_i := \frac{1}{2}\epsilon_{ijk}M^{jk}$ and $K_i := M^{0i}$, we can express the Lie algebra of the Poincar\'e group in the following table:
$$
\begin{array}{c|cccc}
& H & P_j & J_j & K_j \\
\hline\\
H & 0 & 0 & 0 &  P_j \\ \\
P_i & 0 & 0 &  \epsilon_{ijk}P_k & \delta_{ij} H\\ \\
J_i & 0 &  \epsilon_{ijk}P_k &  \epsilon_{ijk}J_k &  \epsilon_{ijk}K_k\\ \\
K_i & -P_i  & -\delta_{ij}H  & \epsilon_{ijk} K_k & -\epsilon_{ijk} J_k \\
&
\end{array}
$$
The table is to be read as follows: the entry whose row is named by the quantity $A$ and whose column is named by the quantity $B$ gives the commutator $\frac{1}{i\hbar}[A, B]$. The same table also applies in the classical Hamiltonian setting, in which each corresponding entry may be read as the Poisson bracket $\{A, B\}$.

We now introduce an important derived quantity, \emph{the Pauli-Lubanski pseudovector} $W_\kappa$, defined by $W_\kappa := \frac{1}{2}\varepsilon_{\kappa\lambda\mu\nu}M^{\lambda\mu}P^\nu$, or in terms of the space/time split,
\begin{equation}
W_0 := \mathbf{J.P} \ ;
\qquad
\mathbf{W} := H\mathbf{J} - \mathbf{P}\times\mathbf{K}\ .
\end{equation}
This pseudovector (so-called because of its behaviour under parity transformations) is always Minkowski orthogonal to the four-momentum: $W_\mu P^\mu= 0$.
Given the commutation relations in the table above, there are two derived quantities which commute with every element in the algebra---these are the so-called \emph{Casimir invariants}. The first is $P_\mu P^\mu = H^2 - \mathbf{P}^2 =: m^2$, which gives the squared relativistic rest mass of the system. The second is $W_\mu W^\mu$, which, if the system is massive and therefore has a rest frame,  is equal to $W_\mu W^\mu = -m^2\mathbf{J}^2$ in that rest frame. For massive representations, $\mathbf{J}^2$ in the system's own frame yields the squared magnitude of the system's spin, and we have the constraint that  $\mathbf{J}^2 = s(s+1)\hbar^{2}$, where $s$ is an integer or half-integer. In massless representations $W_0$ is proportional to the helicity of the system.

The Casimir invariants are special, since in any irrep they take the same value for all states, and so behave in each irrep as a multiple of the identity. They are therefore  interpreted as properties that the system has according to every inertial frame (under a passive construal of the Poincar\'e transformations), or necessarily (under an active construal). With something of a leap of logic, they are often interpreted as  \emph{intrinsic} properties of the system---that is, properties that the system has independently of anything else (e.g.~Castellani 2002). We therefore yield the result that  particles are characterised by their rest mass, and spin (if massive) or helicity (if massless).
At this point we might also bring in the `internal' symmetries, which come with their own Casimir invariants. We thereby obtain the result that  particles are characterised not only by their rest mass and spin/helicity, but also by their  electroweak isospin, electroweak hypercharge and colour charge. (Electric charge is a derivative charge, defined in terms of electroweak isopsin and hypercharge.)

So much for my brief overview of Wigner's identification and its conceptual underpinnings. I might also mention here that an analogue of Wigner's identification may be run for Galilei spacetime. According to this identification, a particle on Galilei spacetime `is' an projective unitary representation of the Galilei group. In this case, projective unitary representations of the Galilei group are equivalent to unitary representations of a central extension of the Galilei group, known as the \emph{Bargmann group.}  The Casimir invariants obtained in this case are mass $M$, minimum energy $E_0 := H - \mathbf{P}^2/2M$ and spin $s$. (See Levy-Leblond 1963 for more.)

So to summarise, Wigner's identification is motivated by the following  ideas: (i) to characterise a system is to characterise its space of possible states, perhaps also to specify the algebra of its quantities; (ii) this space of possible states, and associated algebra, can be thought of as determined by some Lie group, via a representation or realization of some appropriate kind (projective unitary representations in the case of a quantum theory, canonical realizations in the case of a classical Hamiltonian theory); (iii) in the specific case of particles, the relevant Lie group is the automorphism group of the spacetime in which the particle lives. (i)  is ingenious, and will not be questioned in this paper; (ii) is established mathematics; (iii) is the claim I wish to challenge.

\subsection{Where is Wigner's identification used?} \label{WignerUses}

As I said above, particles are not fundamental. There are three regimes in which it appears to be uncontroversial to say that particle-talk is legitimate, if only in the sense of an effective theory: (i) for free field theories, in which interactions are ``turned off''; (ii) in the 1-particle sector of the quantum field's Hilbert space; and (iii) for asymptotic states in particle collisions. (Note that in all three of these regimes the particles in question are either free or approximately free; I return to this below.) In all three of these regimes Wigner's identification delivers the results we expect: let me take them in turn.

First, the free field `particle picture', given in terms of Fock space and creation and annihilation operators.  For simplicity, I will consider the case of the bosonic scalar field. The particle picture for this field is one of two---or perhaps three---complementary conceptual nets which one may cast over the quantum field.  (Here I shall borrow heavily from Baez, Segal \& Zhou 1992.) The second picture is the `wave' or `field picture', which is articulated in terms of field operators and their eigenstates, in which the states of the quantum field are thought of as complex-valued functionals over classical field configurations. A possible third picture Baez \emph{et al} call the `complex wave picture', and is articulated in terms of `eigenstates' of the annihilation operators, which are thought of as antientire functions on the 1-particle Hilbert space---this picture is important in considering the classical limit of bosonic fields.

In the field picture, the central quantities are the field operators $\Phi(z)$, more properly thought of as operator-valued distributions, which map from states in the classical Hamiltonian theory of the free field to  operators in some representation of the Weyl algebra. In infinitesimal form, these are the familiar CCRs:
\begin{equation}
[\Phi(z), \Phi(z')] = i\hbar\Omega(z, z')\ .
\end{equation}
Here $\Omega$ is a the symplectic product defined on the classical phase space $S$, which, because it is a vector space, allows one to bring the symplectic form `downstairs' from the tangent spaces to the phase space itself, thus turning it into a symplectic vector space. It is perhaps more familiar to represent the field operators in terms of localised field configuration and momentum operator-valued distributions. Thus, if $S\ni z = (f, g)$, where $f : R\to\mathbb{R}$ is some classical field configuration on the 3-space $R$ and $g : R\to\mathbb{R}$ is some classical field momentum configuration on $R$, then 
\begin{equation}\label{fieldop}
\Phi(z) = \int_R \textrm{d}^3\mathbf{x}\ \left( f(\mathbf{x})\hat{\pi}(\mathbf{x}) -  g(\mathbf{x})\hat{\phi}(\mathbf{x})\right)\ ,
\end{equation}
where $\hat{\phi}(\mathbf{x})$ is the familiar local field operator, and  $\hat{\pi}(\mathbf{x})$ is its associated conjugate momentum field operator.\footnote{The symplectic product is then given by $\Omega((f,g),(f',g')) = \int_R \textrm{d}^3\mathbf{x}\ \left( f(\mathbf{x})g'(\mathbf{x}) -  f'(\mathbf{x})g(\mathbf{x})\right)$.}

In the particle picture, the absolutely central operators are the creation and annihilation operators $a(\psi), a^+(\psi)$, which are properly thought of as maps from states in the 1-particle Hilbert space $\mathcal{H}_1$ to operators in some representation of the Weyl algebra. In terms of creation and annihilation operators, the CCRs may be written as follows:
\begin{equation}\label{CCRs2}
[a(\psi), a(\varphi)] = 0\ ; \qquad
[a(\psi), a^+(\varphi)] = \langle\psi, \varphi\rangle \ 
\end{equation}
(where the inner product on the RHS of the second equation is on the 1-particle Hilbert space).
The representation of interest for us here is the Fock representation, generated from the vacuum state $|0\rangle$, defined by the condition $a(\psi)|0\rangle = 0$ for all $\psi\in\mathcal{H}_1$, by arbitrary finite applications of  creation operators. The central quantities in this representation are the occupation number operators $N(\psi) = a^+(\psi)a(\psi)$, which, for any normalised state $\psi$ has as its spectrum the natural numbers $\mathbb{N}$.

Crucially, the 1-particle Hilbert space $\mathcal{H}_1$ and classical symplectic vector space $S$ are intimately connected. In essence, $S$ is turned into $\mathcal{H}_1$ by defining on it a complex structure $J$, determined by the free dynamics, which then allows one to define an inner product and close the space under the induced norm.  Specifically, there is a real-linear embedding $K:S\to\mathcal{H}_1$ such that $KJz = iKz$ for all $z\in S$. Classical dynamical evolution on $S$ due the classical Hamiltonian
\begin{equation}
H(z) = H(f, g) = \int\textrm{d}^3\mathbf{x}\ \left(\frac{1}{2}g(\mathbf{x})^2 + \frac{1}{2}f(\mathbf{x})\omega^2f(\mathbf{x})\right)
\end{equation}
where $\omega$ is a linear operator (in the case of the Klein-Gordon field, $\omega^{2} = -\nabla^{2} + m^2$), is then mapped under $K$ to unitary evolution due to the quantum Hamiltonian $\omega$. (In this case, the complex structure $J$ takes the form $J(f, g) = (-\omega^{-1}g, \omega f)$.)

But not only is there an intimate connection between $S$ and $\mathcal{H}_1$; there is also an intimate connection between the field operators and creation and annihilation operators. They are inter-definable, according to the following identities:\footnote{One often finds in the literature, instead of these identities, something along the lines of $\Phi(z) = a(z) + a^+(z)$. This is sloppy, since the field operators are strictly a function on the symplectic vector space $S$ and the creation and annihilation operators are strictly a function on the 1-particle Hilbert space $\mathcal{H}_1$. So a better identity would be $\Phi(z) = a(Kz) + a^+(Kz)$. Ignoring the missing factor of $\hbar$, this identity works perfectly well; but the one presented in the main text is to be preferred, since it entails the familiar expression for the local field operators $\hat{\phi}(\mathbf{x})$ in terms of the momentum creation and annihilation operators $a(\mathbf{k}), a^+(\mathbf{k})$ found in most textbooks.}
\begin{equation}\label{FPduality}
a(Kz) = \frac{1}{2\hbar}\left(i\Phi(z) - \Phi(Jz)\right)\ ;
\qquad
\Phi(z) = -i\hbar\left(a(Kz) - a^+(Kz)\right)\ .
\end{equation}

Before we get back to the Poincar\'e group, let me say something briefly about field-particle duality, since it will be important later (section \ref{GeoInn}). The duality can be articulated precisely as a unitary equivalence between two representations of the quantum field (Baez \emph{et al} (1992, p.~49; Theorem 1.10)). As stated above, the particle representation is given by a Fock space, generated from the vacuum by arbitrary finite applications of creation operators. The field representation is given by some Hilbert space of square-integrable complex-valued functionals $\Psi(f)$ on classical field configurations $f$, in which the local field operators $\hat{\phi}(\mathbf{x})$ act on the state as multiplication by the classical field value $f(\mathbf{x})$, and the local momentum operators  $\hat{\pi}(\mathbf{x})$ act as functional derivative by $f(\mathbf{x})$ (modulo an extra term due to the integration measure). But since the space of classical field configurations is infinite-dimensional, there is no uniquely natural measure on it, and therefore no uniquely natural  Hilbert space. If we choose the measure $\mu$ such that $\textrm{d}\mu(f) \sim \exp(-\frac{1}{4\hbar}\int\textrm{d}^3\mathbf{x}\ f(\mathbf{x})\omega f(\mathbf{x}))$, an instance of a so-called \emph{isonormal} measure, then the resulting representation is unitarily equivalent to the particle representation.

This duality applies only to the free quantum field, since interacting fields have no Fock representation (Earman \& Fraser 2006, Fraser 2008). However, an important fact that seems not to be better appreciated is that the identities (\ref{FPduality}) hold in every representation---or rather, they can be made to hold \emph{by definition} of $a(\psi), a^+(\psi)$ in every representation. Since the field operators $\Phi(z)$ are defined in every representation, it follows that so too must be the creation and annihilation operators $a(\psi), a^+(\psi)$. So too, then,  are the occupation number operators $N(\psi) = a^+(\psi)a(\psi)$. (What may \emph{not} be defined the \emph{sum} of  $N(\psi_i)$ over some complete orthonormal basis $\{\psi_i\}$ of $\mathcal{H}_1$---this, were it defined, would be the total number operator.) And the CCRs (\ref{CCRs2}) are sufficient to establish the following operator identity:
\begin{equation}
(N(\psi)+1)a(\psi) = a(\psi)N(\psi) 
\end{equation}
which is sufficient to show that the spectrum of each $N(\psi)$ is $\mathbb{N}$ in every representation.

All of this goes to show that some kind of particle notion appears to survive even outside of Fock representations. Indeed: the Fock representation appears special now only in virtue of having a well-defined total number operator. I say all of this because one may expect from an adequate characterisation of particles that it applies not only to Fock representations, but in {any} representation in which the creation and annihilation operators are well defined; and as we have seen, that is the case for \emph{all} representations of the CCRs.

What I want to highlight here is that what is crucial to field-particle duality---both in the sense of the unitary equivalence between the field and particle pictures in the free case, and in the more general sense of the identities (\ref{FPduality})---is that the classical field configurations $f(\mathbf{x}), g(\mathbf{x})$, and their quantum counterparts the local field operators $\hat{\phi}(\mathbf{x}), \hat{\pi}(\mathbf{x})$, be defined over the same space as the 1-particle wavefunctions $\psi$ in the domain of the creation and annihilation operators $a(\psi), a^+(\psi)$. If this were to fail, then one could not construe states of the quantum field, thought of as $\Psi(f)$, in terms of particle excitations, e.g.~states in some Fock space (when the field is free). If field-particle duality, at least in the case of free fields, is to be taken as a non-negotiable aspect of the particle concept, then, as we shall see in section \ref{GeoInn}, problems appear to arise for generalisations of Wigner's identification to other spacetimes.

But returning to our overview of the successes of Wigner's identification: how does Wigner's identification fit into the discussion of the free quantum field, above? When the field is free, and we are in  the Fock representation, every single 1-particle Hilbert space in the Fock construction carries the same kind of irrep of the Poincar\'e group. As the field evolves, the unitary dynamics restricted to each 1-particle Hilbert space respects the energy-momentum relation $H^{2} - \mathbf{P}^{2} = m^{2}$ (for the same value of $m$), where $H$ here is the 1-particle Hamiltonian. The quantity $H^{2} - \mathbf{P}^{2}$ (recall) is one of the Poincar\'e algebra's Casimir invariants. Thus our temptation to consider each Fock-space excitation as a particle (of the same kind as all the others) is vindicated by Wigner's identification.

But given the foregoing, this immediately raises a question. When the field is not free, the states $\psi$ in the domain of the creation and annihilation operators $a(\psi), a^+(\psi)$---recall that these operators are still well defined when the field is not free---could not transform under an irrep of the Poincar\'e group. For if they did, then the dynamical evolution of every 1-particle Hilbert space would be accounted for, and the resulting field as a whole would evolve freely, contrary to hypothesis. So how are we to understand the 1-particle states $\psi$ in non-Fock (or inequivalent Fock) representations?

I think a satisfactory answer to this question can be given, on the basis of my rival proposal in section \ref{Section3}. To anticipate, the 1-particle states $\psi$, and their associated algebra of quantities, are better thought of in \emph{all} representations as exactly that: instantaneous states, in which nothing is presumed about their dynamical evolution (and likewise the associated algebra of quantities makes no presumption about dynamical evolution). In that case, these states can be understood in a way that is completely agnostic as to the dynamics. But this line is not compatible with Wigner's identification. This speaks to the problem of dynamical innocence, which I address in section \ref{DynInn}, so let me suspend the current discussion for now.

Returning to the successes of Wigner's identification, it is clear that the identification gives the expected results for the 1-particle sector of the quantum field. That is: this sector contains states of a single particle, alone in the universe. This sector is just the 1-particle Hilbert space $\mathcal{H}_1$ on which  the Fock space is constructed, and we have already seen that this Hilbert space carries an irrep of the Poincar\'e group.
 
 Finally, the asymptotic states. A complete treatment of these ought to go into the details of Haag-Ruelle scattering theory and the assumptions feeding into the derivation of the LSZ reduction formula. I will just offer the briefest of comments. (Very useful detailed discussions are given by Bain (2000), Wallace (2001) and Duncan (2012, \S\S9.3-9.4).) The essence is this: far away from a collision, the states of the interacting quantum field `look like' states of the free quantum field, the particle states of which transform under irreps of the Poincar\'e group. (In the case of Haag-Ruelle scattering theory, we have the \emph{Haag asymptotic theorem}, which is a claim of strong convergence to such free states; see Duncan (2012, pp.~276-7).) Thus Wigner's identification vindicates our expectations that such asymptotic states comprise particles.

To close off this discussion of the successes of Wigner's identification, I will make a brief mention of the irreducibility requirement. It is sometimes claimed that the irreducibility of the rep is to do with the associated particle being elementary, in the sense of non-composite. This is a misunderstanding. Protons, atoms of hydrogen in their ground state, centres of mass, and many more composite systems besides are all associated with irreducible reps. Indeed, even the humble electron, which is surely associated with an irrep, is not truly non-composite: a realistic electron is not just a quantum of the electron-positron field, but some unholy hodgepodge involving \emph{all} fields found in the Standard Model. In a motto: real particles are quasi-particles. 

The proposed link between irreducibility and elementarity is mistaken, but it is close to the truth. The true link is between irreducibility and ignored, or factored-out, internal structure. A proton is associated with a rep that is irreducible because, if you wiggle its internal structure, it ceases to be a proton. An atom of hydrogen in its ground state is associated with a rep that is irreducible because, if you wiggle its internal structure, it ceases to be in its ground state, or ceases to be a bound state (and so a hydrogen atom) at all. A centre of mass is associated with an irrep because centres of mass are by construction divorced from any internal structure. And so on. Therefore our intended target here is not merely a characterisation of elementary particles; it is a characterisation of any particle---indeed any \emph{system}---whose internal structure is, for reasons of classification or perhaps simply convenience, being factored out of consideration. Nevertheless, I will continue to talk in terms of particles.

\subsection{What is the argument?} \label{WignerArgues}

\subsubsection{Wigner's theorem}

I turn now to Wigner's argument for the identification of particles with irreps of the Poincar\'e group. I wish to mark an important distinction between Wigner's \emph{theorem} and Wigner's \emph{argument}. Very briefly, Wigner's \emph{theorem} says that two representations agreeing on transition probabilities are connected by either a unitary or antiunitary. Wigner's \emph{argument} is to the effect that a particle's Hilbert space and associated algebra of quantities is an irreducible, projective unitary  representation of the Poincar\'e group. Wigner's theorem plays a crucial role in this argument, but it is other aspects of the argument that will concern us here.

Wigner's Theorem was first given in Wigner (1931; 1959, pp.~233--6). A more complete proof is in Weinberg (1995, p.~50; 91--96). I will state it in slightly more generality than is found in Weinberg:

\emph{Let $\mathscr{H}$ be the Hilbert space of states for some observer $O$ and $\mathscr{H}'$ be the Hilbert space of states for some observer $O'$, and let $[\Phi], [\Psi]\in\mathscr{H}$ be arbitrary rays for $O$ and let $[\Phi'], [\Psi']\in\mathscr{H}'$ be rays representing those same states, respectively, but as described for $O'$, so that transition probabilities are the same for both observers:}
\begin{equation}\label{WignerA}
|\langle\Phi, \Psi\rangle_O|^2 = |\langle\Phi', \Psi'\rangle_{O'}|^2 .
\end{equation}
\emph{($\langle \cdot, \cdot\rangle_O$ is the inner product defined on $\mathscr{H}$ and $\langle \cdot, \cdot\rangle_{O'}$ is the inner product defined on $\mathscr{H}'$.)
Then there is some operator $U:\mathscr{H}\to\mathscr{H}'$, which is either linear and unitary or antilinear and antiunitary, such that $U\Phi = \Phi'$ and $U\Psi = \Psi'$.
}

The argument continues: Let $\Lambda$ be the transformation which sends $O$ to $O'$; then $U = U(\Lambda)$; i.e.~$U$ is not a function of $O$ or $O'$, but merely of the transformation which connects them. It follows that the Hilbert spaces of any two other observers similarly related by $\Lambda$ are connected by the same operator $U(\Lambda)$. Considering now three observers, connected by the transformations $\Lambda_1:O\to O'; \Lambda_2:O'\to O''$; then we must have that
\begin{equation}
U(\Lambda_2)U(\Lambda_1)=\omega(\Lambda_1, \Lambda_2) U(\Lambda_2\circ\Lambda_1) \ ,
\end{equation}
where $|\omega(\Lambda_1, \Lambda_2)| = 1$. Note: if the representation is not irreducible, then $\omega(\Lambda_1, \Lambda_2)$ is  a matrix which acts a multiple of the identity on each irrep.

Weinberg assumes that $\mathscr{H} = \mathscr{H}'$, but there is no need to make this assumption, and in fact it is not entirely clear why we may assume it.  This theorem is the entire motivation for the search for projective unitary representations in the case where the symmetries all belong to the same connected component of a Lie group. (In this case we can rule out antiunitary representations by considerations of continuity at the identity.) But we must have $\mathscr{H} = \mathscr{H}'$ to get all this off the ground.

\subsubsection{Wigner's argument}

Here is Wigner (1939, pp.~149-50) (I have slightly updated his notation to use angle brackets):
\begin{quote}
The square of the modulus of the unitary scalar product $\langle \psi, \varphi\rangle$ of two normalized wave functions $\psi$ and $\varphi$ is called the transition probability from the state $\psi$ into $\varphi$, or conversely. This is supposed to give the probability that an experiment performed on a system in the state $\varphi$, to see whether or not the state is $\psi$, gives the result that it is $\psi$. If there are two or more different experiments to decide this (e.g., essentially the same experiment, performed at different times) they are all supposed to give the same result, i.e., the transition probability has an invariant physical sense.

The wave functions form a description of the physical state, not an invariant however, since the same state will be described in different co\"ordinate systems by different wave functions. In order to put this into evidence, we shall affix an index to our wave functions, denoting the Lorentz frame of reference for which the wave function is given. Thus $\varphi_l$ and $\varphi_{l'}$ represent the same state, but they are different functions. The first is the wave function of the state in the co\"ordinate system $l$, the second in the co\"ordinate system $l'$. If $\varphi_l = \psi_{l'}$, the state $\varphi$ behaves in the co\"ordinate system $l$ exactly as $\psi$ behaves in the co\"ordinate system $l'$. If $\varphi_l$ is given, all $\varphi_{l'}$ are determined up to a constant factor. Because of the invariance of the transition probability we have
$$
(1)\qquad\qquad |\langle\varphi_l, \psi_l\rangle|^2 = |\langle\varphi_{l'}, \psi_{l'}\rangle|^2 
$$
\end{quote}
Here  Wigner immediately turns to a version of what I have been calling Wigner's theorem. My interest here is how Wigner gets to the equality of transition probabilities, the essential antecedent to Wigner's theorem, and what implications that has  for interpreting the resulting irreducible representation of the Poincar\'e group.

What's striking about the argument above is that, on its face, it is a consideration of passive transformations---that is, transformations between different representations of the same physical state. Indeed, the transformations being passive is crucial: for this justifies holding the transition probabilities---having an `invariant physical sense'---fixed. But if we were to take all this at face value, we would be led to conclude that the apparent multiplicity of dimensions in the resulting representation  is nothing but a representational redundancy, since all rays correspond to the same physical state, albeit as represented in different frames.\footnote{There is an implicit assumption here, that if the  mathematical states $\psi$ and $\psi'$ represent the same physical states, then so does any mathematical state in their linear span. This assumption can fail if the mathematical states in question are statistically mixed; but in this case the states are pure.}

It is difficult to see what purpose would be served by a theory whose Hilbert space contains nothing but different representations of the same physical state. Besides, irreps of the Poincar\'e group are simply not used in this way. There is a reasonable relationalist argument to be given, to the effect that only one physical state is available to a single particle that is alone in the universe (alone, that is, not counting the spacetime manifold and Minkowski metric). But this is an exceedingly special case, and a distraction from our main discussion. Our interest here is in finding the right characterisation of particles, such as (we believe) we find them in the real world. Now, as I will argue below, we do so by providing a characterisation that works \emph{whether or not} the particle is alone in the universe. But we certainly cannot do so if we must first assume that it \emph{is} alone. The particles (we believe) we encounter in the real world are certainly not alone.

Another puzzle about the face-value, passive interpretation of Wigner's argument is that there appears to be no good reason to restrict attention of Poincar\'e transformations. If we must give a picture of the world from some frame, why should it be an inertial one? If we are comparing representations of the same state in two different frames, why should they be related by a Poincar\'e transformation? If a complete picture of the world can be given from the point of view of an inertial frame, then it can be given in a non-inertial one too.

I would like to suggest that there is a straightforward way to convert Wigner's face-value argument into an argument for something more plausible. The conversion requires two assumptions: (i) that passive and active transformations (can) have  equivalent mathematical representations; and (ii)  Poincar\'e transformations are  symmetries of Minkowski spacetime. Now, these assumptions are not only true, but so obviously so that this perhaps explains Wigner not making them explicit in his argument.

We begin with a model of Minkowski spacetime, together with a Lorentz frame and a quantum state
$\langle M, \eta_{ab}, \phi, \psi \rangle$. Here, $\phi = \langle x^0, x^1, x^2, x^3\rangle$ is a global coordinate chart, which is our mathematical representation of a single Lorentz frame.  (If you like, you may assume that the coordinate chart is `adapted' to the Minkowski metric in the sense that $\eta^{mn}\mathbf{d}_mx^\mu\mathbf{d}_nx^\nu = \eta^{\mu\nu}$, where $\eta^{\mu\nu} = \mbox{diag}(+1, -1, -1, -1)$. $\psi : M\to\mathbb{C}$ is the quantum state, represented here as a direct function on the spacetime manifold. The composed function $\psi\circ \phi^{-1}$ is then the quantum state expressed as a familiar function of coordinates; it is Wigner's $\psi_l$.

We may now represent a passive transformation as a transformation on the coordinate chart alone. So, choose some diffeomorphism $h: M\to M$, and drag along the chart $\phi$ by its inverse $h^{-1}$ to obtain
$\langle M, \eta_{ab}, (h^{-1})^*\phi, \psi \rangle$ (I am using the inverse $h^{-1}$ merely for convenience later). Now, since we have transformed only the coordinate chart, the physical state $\psi$ is unchanged; yet we obviously have a different induced function on coordinates $\psi\circ ((h^{-1})^*\phi)^{-1}$. This is Wigner's $\psi_{l'}$.

Now instead consider an active transformation on the original model, induced by the original diffeomorphism $h$. Deliberately keeping the Minkowski metric and coordinate chart fixed, we obtain
$\langle M, \eta_{ab}, \phi, h^*\psi \rangle$. If now $h$ is a Poincar\'e transformation, then it is a symmetry of  the Minkowski metric, and so we have $h^*\eta_{ab} = \eta_{ab}$. Our new model may therefore also be written $\langle M, h^*\eta_{ab}, \phi, h^*\psi \rangle$. 
I suggest this state is Wigner's $\varphi_l$, when he says, `[i]f $\varphi_l = \psi_{l'}$, the state $\varphi$ behaves in the co\"ordinate system $l$ exactly as $\psi$ behaves in the co\"ordinate system $l'$.' (This sentence is the only hint in Wigner's original paper that he intended this flitting between passive and active construals of the Poincar\'e transformations.) For, 
 this model is now manifestly isomorphic with the model $\langle M, \eta_{ab}, (h^{-1})^*\phi, \psi \rangle$ resulting from the passive transformation. Crucially, the induced quantum state as a function of coordinates is identical, since $(h^*\psi)\circ\phi^{-1} = \psi\circ h\circ \phi^{-1} = \psi\circ (\phi\circ h^{-1})^{-1} = \psi\circ ((h^{-1})^*\phi)^{-1}$. Thus $\varphi_l = \psi_{l'}$. But since the models are isomorphic,  we expect them to be indistinguishable as regards physics.

This is not an appeal to `Leibniz equivalence', which is the claim that isomorphic models represent the same physical history. The claim is simply that, when $h$ is a Poincar\'e transformation, the model $\langle M, \eta_{ab}, \phi, h^*\psi \rangle$, ostensibly a \emph{different} physical state as represented in the \emph{same} Lorentz frame, is isomorphic, and therefore  indistinguishable as regards physics, to the model $\langle M, \eta_{ab}, (h^{-1})^*\phi, \psi \rangle$, ostensibly the \emph{same} physical state as represented in a \emph{different} Lorentz frame. In other words: we may take a representation of the same physical state in a different Lorentz frame as proxy for a different possible state in the same Lorentz frame. Crucial to establishing this claim is the assumption that the transformation $h$ is a Poincar\'e transformation, and this explains Wigner's restriction to Poincar\'e transformations.

It may be asked, then, why Wigner didn't pursue his argument by explicitly considering active transformations at the outset.  The reason, I suggest, is that the detour through passive transformations provides  a water-tight justification for the preservation of transition probabilities, which then permits the application of Wigner's theorem to obtain projective unitary representations.
In any case, it seems not everyone was happy with the concision of Wigner's original argument. Streater's (1975, \S2.2) version makes explicit mention of the passive/active distinction, claiming that `[t]he active point of view follows from the passive one', with a reference to Wigner (1963) and relying explicitly on the relativity principle. This is also seen in more recent textbook treatments, e.g.~Schweber \& Bethe (1964) and Sternberg (1995, p.~149). The consensus is then surely that the Poincar\'e transformations are to be construed actively, and the ensuing representation is to be construed as giving a space of \emph{different} possible states for the \emph{same} system, in line with our discussion above.

\subsubsection{Initial doubts}

Let me here articulate my doubts regarding  the above argument. This flitting between passive and active construals of the Poincar\'e transformations requires that we consider a particle alone in the universe. Otherwise, an active transformation on a single particle could not produce the sort of global change that is characteristic of passive transformations. (If I transform my position on the world, the entire world, relative to me, changes.) We must then ask whether it is legitimate to characterise a particle only by its possible states \emph{when it is alone.}

I have two misgivings. The first, already mentioned, is that if we really must imagine the particle alone in the universe, then we would seem vulnerable to a relationalist argument to the effect that the Poincar\'e transformations could not be construed actively, since there is simply not enough structure in the world to make sense of them generating new states.
The second misgiving is that this  simply fails to connect with  how Wigner's identification is used in practice---except in the very special case in which we consider the 1-particle sector of the quantum field. Recall (section \ref{WignerUses}) that Wigner's identification finds use in characterising the asymptotic states of particle collisions---in essence telling us what \emph{counts} as a particle collision. A lonely particle does not collide with anything.

Now both these misgivings could be dispelled if the following could be established: what is possible for a particle on its own is also what is possible for a particle when accompanied (perhaps by other particles, or fields, or whatever). There is a sense in which this is obviously false. For, what is possible for any object surely includes the relations it may stand in with other objects; and these possibilities are excluded when we consider the particle alone. But we can put this objection aside, since we have a precise sense of `what is possible' for the system that does not get into relational states of affairs.

What we want is some characterisation of a particle, in terms of its space of possible states and associated algebra of quantities---that is, in terms of a representation---which is \emph{modular}. That is, once we are equipped with such  a representation, we can build whatever representation we like for systems having that particle as a proper constituent by using the familiar means of forming tensor products, and (if need be) restriction to the symmetric or antisymmetric sectors. This representation, therefore, should apply \emph{whether or not} the particle is alone, and specifically whether or not it is interacting with something else.

We shall see in the following section that irreducible representations of the Poincar\'e group do \emph{not} provide this kind of modularity \emph{unless} the particle, when accompanied, is accompanied in the most boring way possible, i.e.~non-interacting. But note that this at least vindicates Wigner's identification, and his argument for it, for the context in which it is most commonly deployed: namely, in considerations of the asymptotic states of particle collisions. Restricted to this context, Wigner's identification and his argument for it, going \emph{via} Wigner's theorem, seems to be in good shape.


\section{Where Wigner's identification fails} \label{Section2}

Wigner's identification is wrong in chiefly two separate ways, or so I shall argue. The first way is that it has as a necessary consequence that  particles are  free. But particles are in fact rarely free, even approximately. More specifically, there are several empirically successful theories which are apparently about interacting particles, and we should seek to vindicate a face-value reading of these theories' implied ontologies, as being about particles. That is what I aim to argue in subsection \ref{DynInn}.

The second way  that Wigner's identification is wrong is that it misleadingly implies some sort of deep connection between the symmetries of spacetime and the generators of the space of states for a particle in that spacetime. I believe there is no such connection, except that they happen to coincide for \emph{free} particles in Galilei or Minkowski spacetimes. That they do coincide for these spacetimes is a mark of how special these spacetimes are, not the symptom of some deep connection between spacetime symmetries and particle representations. That  is what I aim to argue in subsection \ref{GeoInn}.

I end each subsection with some desiderata for an alternative characterisation of particles. Chief among these desiderata are what I call \emph{dynamical innocence} and \emph{geometrical innocence}, and so these name the relevant subsections.

\subsection{Dynamical innocence}\label{DynInn}

Here's a tempting line of thought. As we have seen in section 1, Wigner's identification, and his argument for it, work when either  particles are alone or when they don't interact with anything. This covers, respectively, the 1-particle sector of the quantum field and the asymptotic states in particle collisions (though only for those particles which are \emph{involved} in asymptotic states), which are its two chief applications. However, the particle picture breaks down when interactions are turned on, and so Wigner's identification---indeed \emph{any} potential rival characterisation of particles---simply fails to apply in these cases. Taken together, Wigner's identification appears to be the right characterisation, since it is true when, and only when, it makes sense to talk about particles at all.

There is already a wrinkle in this line of thought: asymptotic states include stable bound states (such a Hydrogen in its ground state), and if we are to consider such bound states as involving particles, then the particles in question are not (even approximately) free, and so Wigner's identification cannot be right for them. Sticking to the tempting line of thought would require giving up the idea that bound states involve particles at all. 

This attempt to save the tempting line of thought is, to say the least, drastic. (Really? An atom of hydrogen isn't made of particles?) But it gets worse. It cannot even be made palatable by including some acknowledgement that, in the non-relativistic limit (say), we recover a particle picture. For, taking the non-relativistic limit of Wigner's identification, particles become characterised not by the Poincar\'e group but by the Bargmann group (the central extension of the Galilei group); and any such characterisation constrains the individual particle Hamiltonians to take the form
\begin{equation}
H = \frac{\mathbf{P}^2}{2M} + \mbox{\emph{const.}}
\end{equation}
where $M$ is the central extension with dimensions of mass,  identified as the mass of the particle. This is the Hamiltonian for a non-relativistic \emph{free} particle, and does not even approximately hold for (say) an electron in the hydrogen ground state. So even by taking the non-relativistic limit we do not  recover particles under Wigner's identification---not even in approximation.

Put this failure alongside the empirical success of theories---effective theories, to be sure---of interacting particles. We have of course the theory of the non-relativistic Hydrogen atom, encapsulated in the Schr\"odinger-Coulomb  equation, but also its relativistic counterpart, encapsulated in the Dirac-Coulomb equation. We also have more complicated theories, such as those connected with the two-body Dirac equation or the Salpeter equation. These theories are woefully understudied in the philosophy of physics literature, as are their connections and---one would hope!---reductions (in some regime of approximation) to quantum field theory. There are number of celebrated results in this direction, starting with Gell-Mann \& Low (1951), and continuing with e.g.~Gorelick \& Grotch (1977), Crater \& Van Alstine (1999) and Lindgren \emph{et al} (2005), but as Weinberg (1995, p.~560) laments, `[i]t must be said that the theory of relativistic effects and radiative corrections in bound states is not yet in satisfactory shape.'

Now each of these theories, though enjoying an astounding degree of empirical success, have their limits. They are not fundamental theories, after all. But particles are not fundamental either. And with the implied ontology of any effective theory we face the same dilemma: elimination or vindication? So we must now ask: ought we to take an eliminativist attitude to the particles apparently described in these effective theories, or ought we seek some way to make sense of the claim that interacting particles are real, though not fundamental, and specifically even the things that these theories are about? Are interacting particles to be thought of like phlogiston, a vestige of a defunct research programme? Or are they, like their free counterparts, to be thought of as real but emergent in certain regimes, the regimes in which the theories ostensibly about them enjoy considerable empirical success? Adherence to Wigner's identification counsels elimination, but I suggest this is a counsel of despair.

Perhaps an alternative to Wigner's identification is possible which both: (i) captures both free and interacting particles in stable bound states---and, we might expect, plenty more besides---and (ii) aligns with Wigner's identification precisely where  it works, namely when particles are not interacting.  Would this not be a preferable characterisation of particles? I suggest that it would. Desideratum (i) might be called \emph{dynamical innocence}, since we seek a characterisation of particles which is innocent as to the dynamics, so that it may apply whatever the dynamics. I aim to present such an alternative in section 3.

\subsection{Geometrical innocence} \label{GeoInn}

If Wigner's identification were right, then one might expect that the proposed link between symmetries of the background spacetime and generators of the particle's state space would generalise. A promising result in this direction concerns another case of typical physical interest, namely Galilei spacetime. In this case, projective unitary representations of the Galilei group are equivalent to unitary representations of a central extension of the Galilei group, known as the Bargmann group. Irreps of this group yield exactly what we seem to  want: a space of wavefunctions on 3-space (possibly with spin) and an associated algebra of quantities comprising mass, energy, momentum, angular momentum and (Galilei) boosts.\footnote{It is worth mentioning that the standard position operator is also an element of this algebra. By defining $\mathbf{Q} = (t\mathbf{P} - \mathbf{C})/M$, where $\mathbf{C}$ is the generator of Galilei boosts, $\mathbf{P}$ is momentum, $M$ is mass and the parameter $t$ is  time, it may be checked that $\mathbf{Q}$ uniquely has all of the properties one expects of a non-relativistic position operator: namely, it obeys the standard Heisenberg CCRs with itself and the momentum, and each of its components commute with Galilei boosts along orthogonal directions. Jordan (1980, \S1) offers a short proof, but note that he defines $\mathbf{Q}$ as just $\mathbf{C}/M$. The difference in minus sign is accounted for by his taking Galilei boosts as ``passive'' while I take them as ``active''. }

However, if the considerations in section \ref{DynInn} are along the right lines, then we don't in fact want this, since here too we are stuck with only free particles. But never mind that: the purpose of this section is to argue a separate point.  Consideration of other spacetimes  show rather vividly that the proposed link between spacetime symmetries and generators of  a single particle's state space (even when we restrict attention to free particles) come unstuck.

Consider first a spacetime with fewer symmetries than Galilei spacetime, say ``full'' Newtonian spacetime. Recall (e.g.~Earman 1989, ch.~2) that such spacetimes can be modelled with the ordered tuple $\langle M, h^{ab}, t_{a},\nabla, \xi^a\rangle$, where $M$ is a smooth manifold, diffeomorphic to $\mathbb{R}^4$, $h^{ab}$ is a Euclidean spatial ``metric'', $t_a$ is a temporal ``metric'', compatible with $h^{ab}$ in the sense that $h^{an}t_n = \mathbf{0}$, $\nabla$ is a flat affine connection compatible with $h^{ab}$ and $t_a$ in the sense that $\nabla_ah^{bc} = \nabla_at_b = \mathbf{0}$, and $\xi^a$ is a unit timelike vector field (so $t_n\xi^n = 1$ everywhere), representing a standard of rest. This spacetime has as symmetries precisely the Galilei symmetries \emph{except} boosts, i.e.~spatial and temporal translations and spatial rotations. This is just the direct product of the Euclidean group in 3 dimensions with the additive group $\langle \mathbb{R}, +\rangle$.

Following a generalisation of Wigner's identification, particles in this spacetime would be expected to be associated with irreducible projective unitary representations of this group. The algebra which generates this group then produces energy, momentum and angular momentum, but no boost generators. But without the boost generators, we cannot define position, and so the resulting `particles' have no associated position operator. While it is commonly accepted that relativistic particles have no adequate position operator (a claim I shall be challenging below), the absence of one for nonrelativistic particles would seem to be unacceptable.

To press the point, consider now Aristotelian spacetime, which can be modelled with the ordered tuple $\langle M, h^{ab}, t_{a},\nabla, \gamma\rangle$, in which the standard of rest has been replaced by a single timelike geodesic $\gamma$, representing the centre of the universe. This spacetime  has as symmetries only temporal translations and spatial rotations (around $\gamma$). In this case, under a generalisation of Wigner's identification, particles have as quantities only energy and angular momentum. The representation is so paltry that its states cannot  even be viewed as wavefunctions (perhaps with spin)  on 3-space: rather, they are (or can be viewed as) wavefunctions on the sphere.

This result isn't just counter-inuitive: it blocks any particle picture for a quantum field defined on this spacetime. For,  field-particle duality requires the particles to have wavefunctions defined over the same spacetime as the field operators. Otherwise we cannot reconstrue states of the quantum field in terms of particle-like excitations. Yet here the field operators would still be functions over the entire Aristotelian spacetime, while the Wigner-style particles have wavefunctions merely on some (time-extruded) sphere.\footnote{More rigorously, the field operators are operator-valued \emph{distributions} over 3-space or spacetime, depending on the specific framework, but the point still stands.}  The corresponding Fock space for the particles is simply not matched to the Hilbert space of quantum field states. (And, just in case you find the appeal to Aristotelian spacetime too recherch\'e, we obtain an identical result for any \emph{relativistic} spacetime with the same symmetry group as Aristotelian spacetime: for example, Schwarzschild spacetime.)

Going even further in this direction, but reverting to relativistic spacetimes, it must of course be admitted that a generic classical spacetime background has no symmetries at all. In a loose but robust sense of `almost all', almost all solutions to the Einstein field equations in general relativity specify a metric field with no Killing vector fields. Certainly the spacetime we live in, if it is to be approximated as a classical general relativistic spacetime, can be expected to have no exact symmetries. This is a \emph{total disaster} for the Wigner-style identification of particles, since the relevant group is the trivial group, and its irreducible representations have no structure at all. Yet clearly there are emergent structures in the quantum field defined over such a spacetime which we would intuitively want to identify as particles: namely, approximately localised excitations which remain stable at least over some salient timescale. 

It does not seem satisfactory to reply to this that the spacetime we live in is approximately flat, and so \emph{via} Wigner's identification we can say that, approximately, there are particles. First of all, our spacetime is not approximately flat in plenty of regions of interest. Secondly, let's drill down into what it could mean to say that, merely approximately, there are particles. I presume it  means: there are things  that it's approximately correct to describe as particles. But what are these `things'? Excitations of the quantum field(s) which may be more or less well localised, which approximately obey dynamical equations and which may persist for some sufficiently long period of time. Could we not call these particles too? Is our concept of particle so brittle?

Going in the other direction, now consider a spacetime with more symmetries than Galilei spacetime. Take for example Leibnizian spacetime, which can be modelled with the ordered tuple  $\langle M, h^{ab}, t_{a}\rangle$, where the `metric' fields $h^{ab}$ and $t_a$ are as for Galilei spacetime. Leibnizian spacetime comes with no affine connection, and so no standard of straightness. As a result, its automorphism group is very large indeed---in fact we have an instance of gauge symmetry here, since arbitrary \emph{time-dependent} translations and rotations will preserve the spacetime structure. (We also have time-translation symmetry, as usual.) The corresponding Wigner-style irreducible representation is, as far as I'm aware, not something that has been studied in detail in the physics literature, but with some extrapolation it seems we are going to end up with something far too rich to preserve field-particle duality. The field operators, as usual, will be functions on spacetime, while the 1-particle wavefunctions will have to be defined over a space rich enough to accommodate momentum, velocity boosts (i.e.~Galilei boosts), acceleration boosts, jerk boosts, and so on. There does not seem to be any obvious way to ``squeeze'' such states into wavefunctions defined over 3-space or spacetime, as would be required to consider particles as excitations in the quantum field.

It may be objected at this point that the relevant link implied by Wigner's identification is not, in fact, between \emph{spacetime} symmetries and the generators of a single particle's state space, but rather between the \emph{dynamical} symmetries of the quantum field's equation of motion and  the generators of a single particle's state space. In the case of Minkowski or Galilean field theory, the spacetime and dynamical symmetries match (as per Earman's (1989, ch.~2) celebrated symmetry principles SP1 and SP2), and so \emph{in these cases} the spacetime symmetries are a good guide. In other spacetimes, the dynamical symmetries governing the quantum field may not match the spacetime symmetries, and in these cases, the objection goes, a Wigner-style identification ought to go \emph{via} the dynamical symmetries of the field, not the symmetries of the background spacetime structure.

There is something to this suggestion. Restricted to the 1-particle sector of the quantum field's Fock space, the dynamical symmetries of the 1-particle state space just \emph{are} the dynamical symmetries of the field as a whole, and so we should expect the symmetries to match. It even stands to reason that non-interacting particle states, such as asymptotic states in particle collisions, should obey  the same symmetry group, since dynamically speaking each free particle is as good as alone in the universe.

However, this suggestion cannot, I think, ultimately work. There are two issues. The first is that moving to the dynamical symmetries governing the field may still lead to exactly the same problem as I articulated above: namely, a breakdown of field-particle duality. For the quantum field's equation of motion may have no non-trivial dynamical symmetries at all, and yet the field operators are still be defined over the full spacetime. In this case again, a (modified) Wigner-style identification leads to trivial particle representations---that is, 1-particle wavefunctions defined only at a point (but which point?), so that once again the particles cannot be construed as excitations in the quantum field, which is spread across all of spacetime.
So the first issue is that the proposed amendment to Wigner's identification does not solve the original problem.

The second issue is that the proposed amendment still focusses on the wrong thing: symmetries.
Recall that what is wanted from a characterisation of a particle is a complete specification of that particle's space of possible states---i.e.~everything that is possible for the particle, independent of whether the particle is alone or accompanied---which is generated by some associated algebra of quantities. We therefore demand that the algebra of quantities  exhaustively generate the state space. What right have we to expect that the transformations generated by these quantities are dynamical symmetries? That is, why should \emph{any} two genuinely physically distinct states in the particle's state space enjoy some kind of equivalence with respect to the dynamics? (Of course, the modified Wigner identification demands even more: it demands that the full state space be spanned entirely by states, any two of which are equivalent with respect to the dynamics.) We certainly do not expect anything like this for the quantum field as a whole: the field's equation of motion may possess little or no dynamical symmetries, and yet its state space may be incredibly rich. 

It seems to me that these considerations suffice to show that, when it comes to characterising a particle by its space of possible states, a focus on symmetries---whether spacetime or dynamical---is simply wrong-headed. What we want is a specification of everything that is possible for a particle, \emph{whether or not} two or more of these possibilities possess some kind of equivalence. What the above suggestion seems to get right is that the transformation group which generates the single particle state space---at least when the particle is free---should also generate, though under a different representation, distinct states for the entire quantum field. And while we may expect the representation of this group to be irreducible in the case of a single particle---since these transformations exhaust what can be done to the particle---it should be reducible in the case of the entire field. What seems  to be irrelevant in any of this is whether or not such transformations are \emph{symmetries}.

Minkowski spacetime and Galilei spacetime are special: they are highly symmetric, both having a 10-dimensional automorphism group. The size of these groups seems just enough to generate the full gamut of possible states for a free particle, and no more. However,  as we have seen, less symmetric spacetimes are associated with a more impoverished algebra of generators, and a correspondingly implausibly impoverished space of states for (even free) particles. Meanwhile more symmetric spacetimes are associated with an algebra of generators that is far too generous, and a correspondingly implausibly generous space of states for (even free) particles.  

What is still unclear is this: if the guiding idea behind Wigner's identification, or some modification of it along the lines we have been considering, is wrong, then why does it work so well for free particles in a Minkowski or Galilean spacetime? I hope to give an answer to this in the following section, where I lay out an alternative characterisation. To summarise what is sought from such an alternative characterisation as regards spacetime geometry, we want  it to: (i) capture particles in spacetimes other than Minkowski (or Galilean), in such a way that particle-field duality is preserved (at least in the non-interacting case); and (ii) align with Wigner's identification precisely where, as we have seen, it works, namely when the spacetime is (at least approximately) Minkowskian or Galilei. We might add that it also must (iii) explain what is special about Minkowski and Galilei spacetimes, such that Wigner's identification works there (at least for free particles). 

We might call desideratum (i) \emph{geometrical innocence}, since what is sought is a characterisation of particles which is not tied to the spacetime having the apparently very special form that we see in Minkowski and Galilei spacetimes. There is perhaps no reason to think that such a characterisation need be completely agnostic as to metrical structure---perhaps, like the generalised Wigner-style identification, the particular spacetime is relevant as input to any particular characterisation, but the function from spacetimes to representations nevertheless takes the same form. But in fact, as I will hope to show, the rival characterisation I wish to suggest \emph{is} agnostic in this stronger sense: no background metrical structure needs to be assumed at all.

\section{What's the alternative?}\label{Section3}

\subsection{The QPS algebra}

So we seek some alternative characterisation of particles which is both dynamically and geometrically innocent, preserves field-particle duality (at least in the case of non-interacting particles), and which captures---or better, explains---the success of Wigner's identification in the case of free particles on Minkowski spacetime. The fact that dynamical innocence is sought suggests that it would be a mistake to put any constraints on the characterisation as regards translations in time. Yet we know from the success of Wigner's identification, and its analogue for Galilei spacetime, that the 10-dimensional group offers the right kind of richness in generating the particle's space of states. Dropping time translations, then, it seems sensible to look for some 9-dimensional Lie group.

Such a Lie group is already at hand in the non-relativistic setting. For in that setting there is a characterisation of particles that is perhaps even more familiar than the one going \emph{via} the Bargmann group. That characterisation involves the Heisenberg group, and its associated generators position $\mathbf{Q}$ and momentum $\mathbf{P}$, and the spin group, and its associated generators the spin operators $\mathbf{S}$:
\begin{equation}\label{QPS}
[Q^i, Q^j] = [P_i, P_j] = [Q^i, S^j] = [P_i, S^j] = 0\ ; \qquad
[Q^i, P_j] = i\hbar \delta^i_j\ ; \qquad [S^i, S^j] = i\hbar \epsilon^{ijk}S^k\ .
\end{equation}
Together, these 9 quantities generate a 9-dimensional group. Moreover, their associated irreducible representations appear to be both dynamically and geometrically innocent. Dynamically innocent, because no Hamiltonian is included in the generating algebra, and as we know, the commutation relations above hold for a wide variety of potentials. Geometrically innocent, since no non-trivial geometric structure is required to specify the commutation relations. (It is crucial for this that the components of position $Q^i$ are construed as contravariant, and the components of momentum $P_i$ are construed as covariant, under a point transformation.) For the rest of this section I will call the algebra articulated in (\ref{QPS}) above the \emph{QPS algebra.}

With the QPS algebra, we can build the 10 generators of the Bargmann group. The momentum $\mathbf{P}$ appears already as primitive, and we also have:
\begin{equation}\label{BargmannAlgebra}
\begin{array}{rcl}
H & = &\displaystyle \frac{\mathbf{P}^2}{2M} + E_0 \ ;
\\\\
\mathbf{J} &=& \displaystyle\mathbf{Q}\times\mathbf{P} + \mathbf{S}\ ;
\\\\
\mathbf{C} &=&\displaystyle t\mathbf{P} - M\mathbf{Q}\ ;
\end{array}
\end{equation}
where $H$ is the energy, $\mathbf{J}$ is the angular momentum, $\mathbf{C}$ is the (``active'') Galilei boost operator and $M$ and $E_0$ are constants. With these specifications, it may be checked that the commutation relations of the Bargmann algebra are satisfied.\footnote{The explicit dependence of $\mathbf{C}$ on $t$ is often suppressed, but this term is required to obtain the correct commutation relations between components of $\mathbf{Q}$ and of $\mathbf{C}$, namely $[Q^i, C_j] = i\hbar \delta^i_j t$. Hence the 3-parameter family of unitaries $U(\mathbf{v}) = \exp(-\frac{i}{\hbar}\mathbf{v.C})$ generates the transformations $U^\dag(\mathbf{v})Q^i U(\mathbf{v}) = Q^i + v^it$.} 

In fact we may see the specifications of $H$ and $\mathbf{C}$ as a kind of loss of dynamical and geometrical innocence. We lose dynamical innocence in specifying a particular form for the generator of time translations $H$, and we lose geometrical innocence in the specifying both $H$ and $\mathbf{C}$, since the dot product of $\mathbf{P}$ in $H$ implies a spatial Euclidean metric, and the specific form of  $\mathbf{C}$ implies Galilei spacetime structure, and requires a spatial Euclidean metric to convert the position to a covariantly transforming vector. In this way, we recover a Wigner-style identification as a special case issuing from a specific form of the Hamiltonian.

But since the QPS algebra is both dynamically and geometrically innocent, what is stopping us applying it in the relativistic setting too? As far as I am aware, this idea was first had by Foldy (1956). Foldy's motivation was different to mine: he was interested in unifying the relativistic 1-particle equations---Klein-Gordon, Dirac, Proca, etc.---into a common framework which revealed their deep similarities. The common---Foldy calls it \emph{canonical}---form of these equations turns out to be (Foldy 1956, p.~573)
\begin{equation}
i\hbar\partial_t\psi = \Lambda\omega\psi\ ,
\end{equation}
where $\psi$ is a $2(2s+1)$-component wavefunction defined over 3-space, intended to capture both particle and anti-particle states, 
$\omega = \sqrt{-\hbar^{2}\nabla^2 + m^2}$, and $\Lambda$ acts as the identity on $2s+1$ of the components of $\psi$ and as $-1$ on the remaining $2s+1$ components. 

Instrumental in reaching this unifying equation is the construction of the Poincar\'e generators in terms of the QPS algebra. Specifically, assuming the momentum $\mathbf{P}$ as already primitive, 
\begin{equation}
\begin{array}{rcl}
H & = &\displaystyle \Lambda\omega\ ;
\\\\
\mathbf{J} &=& \displaystyle\mathbf{Q}\times\mathbf{P} + \mathbf{S}\ ;
\\\\
\mathbf{K} &=&\displaystyle t\mathbf{P} - \mathbf{Q}\cdot H + \frac{\Lambda \mathbf{S}\times\mathbf{P}}{\omega + m}\ ;
\end{array}
\end{equation}
where $\omega := \sqrt{\mathbf{P}^2 + m^2}$,  $\Lambda$ acts as +1 (respectively, -1) on positive- (respectively, negative-) frequency states, and the product $\mathbf{Q}\cdot H := \frac{1}{2}(\mathbf{Q}H + H\mathbf{Q})$. (Note: my Lorentz boost generators $\mathbf{K}$ differ from Foldy's by a minus sign; this is because I take Lorentz boosts $\mathbf{K}$ ``actively'' as opposed to ``passively''.)

The third term in the expression for the Lorentz boost operator $\mathbf{K}$ may seem surprising, if not hideous, but it can be motivated. Consider a positive-frequency massive particle in its own rest frame. In this rest frame, we would expect the particle's Pauli-Lubanski pseudovector to be given, in terms of the QPS algebra, by $W_\mu = (0, m\mathbf{S})$, since the particle's momentum vanishes in its own frame. Now boost this vector in the lab frame with an active Lorentz boost with the velocity $\mathbf{P}/H$ (i.e., the velocity of the particle in the lab frame). Such a Lorentz boost is given by the matrix
\begin{equation}
\frac{1}{m}\left(
\begin{array}{cc}
H &  -\ \mathbf{P}\rightarrow \\\\
|   &   \\
\mathbf{P}    &    m + (H - m)\hat{\mathbf{P}}\hat{\mathbf{P}}^T\\
\downarrow  
\end{array}
\right)
\ =\
\frac{1}{m}\left(
\begin{array}{cc}
H &  -\ \mathbf{P}\rightarrow \\\\
|   &   \\
\mathbf{P}    &   \displaystyle m + \frac{{\mathbf{P}}{\mathbf{P}}^T}{H+ m}\\
\downarrow  
\end{array}
\right)
\end{equation}
Applying this to $W_\mu$ we obtain, in the lab frame,
\begin{eqnarray}
W_0 &=& \mathbf{S.P}
\\
\mathbf{W} &=& m\mathbf{S} + \frac{(\mathbf{S}.\mathbf{P})\mathbf{P}}{H+ m}
\\
 &=& H\mathbf{S} - (H -m)\mathbf{S} + \frac{(\mathbf{S}.\mathbf{P})\mathbf{P}}{H+ m}
\\
&=& H\mathbf{S} - \mathbf{P}\times\frac{\mathbf{S}\times\mathbf{P}}{H+m}
\label{PLspin}
\end{eqnarray}
But we already have that $\mathbf{W}
= H\mathbf{J} - \mathbf{P}\times\mathbf{K}$, this just is the definition of the spatial part of the Pauli-Lubanski pseudovector. If we assume the usual form $\mathbf{J} = \mathbf{Q}\times\mathbf{P} + \mathbf{S}$ and demand that $\mathbf{K} = t\mathbf{P} - \mathbf{Q}\cdot H + \mathbf{N}$, where $\mathbf{N}$ is to be determined, then after some cancelling of terms we derive that
\begin{equation}
\mathbf{P}\times\mathbf{N} = \mathbf{P}\times\frac{\mathbf{S}\times\mathbf{P}}{H+m}\ .
\end{equation}
All we need now assume is that $\mathbf{N}$ is perpendicular to $\mathbf{P}$ to derive that 
\begin{equation}
\mathbf{N} = \frac{\mathbf{S}\times\mathbf{P}}{H+m}\ ,
\end{equation}
as in the Foldy decomposition above.

The Foldy decomposition appears to meet the desiderata we have accrued over the course of this paper. Certainly, the QPS algebra is both dynamically and geometrically innocent, for the reasons stated above. Moreover, this alternative characterisation of particles in terms of the QPS algebra captures Wigner's identification precisely when it should, namely when the relevant Hamiltonian is the free one. That is: with the Foldy decomposition of the Poincar\'e generators as above except for the specification of $H$, it may be checked that the Poincar\'e algebra is obeyed iff the Hamiltonian obeys the energy-momentum relation $H^{2} - \mathbf{P}^{2} = \mbox{\emph{const.}}$ An upshot of this is that the mass of a particle is seen to emerge dynamically, i.e.~when the Hamiltonian is of the right form. This is in contrast to the spin of the particle, which is not determined dynamically, but is as much a necessary property of the particle as it seems under Wigner's identification.

For the remainder of this paper, I intend to provide a justification for the Foldy decomposition, and thereby the characterisation of a particle as an irrep of the QPS algebra, in the form of a theorem. The mechanics of the proof are elementary: they involve essentially only commutation relations. But as far as I know it is novel. The theorem suffices only to recover the QPS algebra for Wigner representations; and my main objections against Wigner's identification have been along the lines that we need a way to vindicate particle-talk outside of Wigner representations. So the following section is a separate plank of my case against Wigner's identification and for a characterisation of the QPS algebra.

\subsection{Spin-orbit decompositions}

In this section I report my main theorem, the proof of which is in an appendix. The idea behind my main theorem is that the generators of the Lorentz group, $\mathbf{J}$ and $\mathbf{K}$, can be decomposed into purely orbital and purely spin components; that there are two senses in which this decomposition could be said to be natural, and that these two senses turn out to be equivalent. The first sense of `natural' is that the purely orbital components of these generators on their own should, with the Hamiltonian $H$ and momentum $\mathbf{P}$, generate a Wigner representation with the same mass but zero spin. The second sense of `natural' is that the primitive quantities in this decomposition---which are position $\mathbf{Q}$, momentum $\mathbf{P}$ and spin $\mathbf{S}$---obey the QPS algebra.  Thus the   QPS algebra is justified on the grounds that it is equivalent to the only sensible decomposition of the Lorentz group generators into purely orbital and spin components. The sense in which this is the only sensible decomposition will be discussed in more detail in section \ref{NWunique}.

The theorem is as follows:
\begin{theorem}
Take any massive Wigner representation $(m, s)$ of the Poincar\'e algebra, with generators $H, \mathbf{P}, \mathbf{J}, \mathbf{K}$. Then there are unique $\mathbf{Q}$ and $\mathbf{S}$ such that:
\begin{equation}\label{orbspindec}
\begin{array}{rclcrcl}
 \mathbf{J} &=& \mathbf{L} + \mathbf{S}\ ,& \mbox{where} & \mathbf{L} &:=& \mathbf{Q} \times\mathbf{P}\ ;
 \\
\mathbf{K} &=& \mathbf{M} + \mathbf{N}\ ,& \mbox{where} & \mathbf{M} &:=& t\mathbf{P} - \mathbf{Q}\cdot H \mbox{ and } \mathbf{N.P} = 0\ ; 
\\
\ [\mathbf{Q}, \Lambda] &=& \mathbf{0}\ , & \mbox{where} & \Lambda &:=& H (H^2)^{-\frac{1}{2}}\ .
\end{array}
\end{equation}
and such that either of  the following claims is true:
\begin{enumerate}
\item [(i)] The quantities $H, \mathbf{P}, \mathbf{L}, \mathbf{M}$ obey the Poincar\'e algebra associated with the representation $(m, 0)$.

\item[(ii)]  $\mathbf{Q}$, $\mathbf{P}$ and $\mathbf{S}$ obey the QPS algebra. That is, $\mathbf{Q}$ and $\mathbf{P}$ obey the  Heisenberg algebra; the components of $\mathbf{S}$ obeys the familiar angular momentum algebra;  $\mathbf{Q}$ and $\mathbf{P}$  commute with $\mathbf{S}$; and  $\mathbf{Q}, \mathbf{P}$  and $\mathbf{S}$ act irreducibly on the positive-frequency and negative-frequency subspaces, respectively.
\end{enumerate}
Assuming (\ref{orbspindec}), if either of (i) or (ii) is true, the other one is too.
Moreover,  the following claims are also true:
\begin{itemize}
\item $\mathbf{L}, \mathbf{S}, \mathbf{M}$ and $\mathbf{N}$ are all separately conserved.
\item $H = \Lambda\omega$, where $\omega := \sqrt{\mathbf{P}^2 + m^2}$ and $\Lambda$ has eigenvalues $\pm 1$.
\item $\mathbf{Q}$ is the Newton-Wigner operator.
\item $\mathbf{S}$ forms the  unique spin algebra associated with $s$; i.e.~$\mathbf{S}^2 = \hbar^2 s(s+1)$.
\item  The generators $H, \mathbf{P}, \mathbf{J}, \mathbf{K}$ take the Foldy form; in particular $\mathbf{N}$ is of the form $\mathbf{N} = \frac{\Lambda\mathbf{S}\times\mathbf{P}}{\omega + m}$.
\item $\mathbf{Q}$ transforms covariantly under $\mathbf{M}$ (but  not $\mathbf{K}$ if $\mathbf{S} \neq\mathbf{0}$).
\end{itemize}
\end{theorem}

There are two key upshots to this theorem, going in opposite directions. The first upshot is that, lurking in any massive Wigner representation, defined in terms of the Poincar\'e generators, we can find a position operator $\mathbf{Q}$ and a spin operator $\mathbf{S}$, such that $\mathbf{Q}$ and $\mathbf{S}$, together with the momentum $\mathbf{P}$, obey the canonical and familiar QPS algebra, and $\mathbf{Q}$ and $\mathbf{P}$ generate what can be called with justification the \emph{orbital} degrees of freedom of the particle, while the components of $\mathbf{S}$ generate the spin degrees of freedom. Going in the other direction, the second upshot---and this is the most significant upshot for my position in this paper---is that we may take $\mathbf{Q}, \mathbf{P}$ and $\mathbf{S}$ as primitive, and derive the Poincar\'e algebra, \emph{so long as we specify} that the Hamiltonian takes the form $\sqrt{\mathbf{P}^2 + m^2}$.

As discussed above, the significance of this that the QPS algebra is plausibly both dynamically and geometrically innocent. Dynamical innocence is easily demonstrated by the fact that the QPS algebra has no Hamiltonian, and is compatible with any Hamiltonian that is some function of position, momentum and spin. Geometrical innocence is easily demonstrated by  constructing generators of symmetry groups for alternative spacetimes. For example, we obtain the Bargmann group by specifying the generators as in (\ref{BargmannAlgebra}).

Crucial also, in terms of our desiderata for the particle concept, was upholding particle-field duality as far as possible outside of the restricted regime in which the field is free (and relativistic). Thanks to the Stone-von Neumann Theorem in the case of position and momentum, and the Gel'fand-Naimark Theorem in the case of spin, representations of the QPS algebra all live in the Hilbert space $L^2(\mathbb{R}^3)\otimes \mathbb{C}^{2s+1}$, which meshes perfectly with local $(2s+1)$-component field operators defined over $\mathbb{R}^3$. This matter is discussed in more detail in section \ref{1field}.

To close off the discussion in this section, we finally  turn to a dangling question concerning Wigner's identification and its non-relativistic counterpart: If a particle is not an irrep of the Poincar\'e group (in the relativistic case) or the Bargmann group (in the non-relativistic case), then why do these identifications work so well when the particle is free? Part of the answer, as we have seen, is that the QPS algebra realises either the Poincar\'e algebra or the Bargmann algebra precisely when the  1-particle Hamiltonian takes the free form $\sqrt{\mathbf{P}^2 + m^2}$ or $\frac{1}{2m}\mathbf{P}^2$, respectively. But a deeper part of the answer is that the Poincar\'e and Bargmann groups are special, from the vantage point of the QPS algebra. Their generators are rich enough to cycle through the totality of  possibilities conceived in terms of position, momentum and spin, via translations, rotations and boosts, and no richer. They have exactly the right amount of richness precisely because of their connection to the Principle of Relativity, which holds that the laws of physics do not choose between inertial frames. It is precisely the transformations between inertial frames---which according to the Principle of Relativity should be symmetries---which take us between distinct possibilities for position, momentum and direction of spin. But, as I emphasised above, symmetry has nothing to do with it: this is all about transforming between states in the full space of  possibilities for a particle, whether or not such transformation are symmetries.

\subsection{Is  Newton-Wigner position uniquely preferred?}\label{NWunique}

The main theorem also shows that $\mathbf{Q}$ is the Newton-Wigner position operator, thus singling it out, as many previous treatments have done (e.g.~Newton \& Wigner 1949, Wightman 1962, Fleming 1965, Jordan 1980, Schwarz \& Guilini 2020), as the only reasonable option. The pathologies of this operator have been much discussed in the philosophy literature (Redhead 1995, Malament 1995, Fleming \& Butterfield 1999, Wallace 2001, Halvorson 2001).  I wish to address two in particular, since I believe these are the most pressing. In this and the following section I hope to make peace with both pathologies.
\begin{itemize}
\item \emph{Failure to transform `covariantly'.} If the spin is non-zero, then the Newton-Wigner position operator does not transform as the spatial part of a Lorentz-covariant four-vector under boosts generated by the full boost operator $\mathbf{K}$. (See Proposition \ref{Lemma2} in the appendix.)

\item \emph{Superluminal propagation and failure of microcausality.} Given any state perfectly localised, according to  Newton-Wigner position, in some region $R$ at some time $t$, and a Hamiltonian bounded from below, at any time $t+\epsilon$, where $\epsilon > 0$, the state has non-zero amplitude arbitrarily far away from $R$. And this is so, despite the fact that the spectrum of the velocity operator is bounded by the speed of light. Relatedly, localised projections $P_R(t)$ and $P_{R'}(t')$, where $R$ at $t$ and $R'$ at $t'\neq t$ are spacelike separated (that is to say, localised in terms of Newton-Wigner position), do not commute: i.e.~$[P_R(t), P_{R'}(t')]\neq 0$ (they \emph{do} commute if $t' = t$). This particular pathology was first discussed by Hegerfeldt (1974), and is the centrepiece of what is now known as Malament's Theorem (Malament 1995).
\end{itemize}
The second pathology deserves its own treatment, which I offer in the following section.

The first pathology was addressed rather early on, by Fleming (1965). In brief, the resolution is that it was always a mistake to assume that the trajectory  of a relativistic particle, understood in terms of Newton-Wigner position, defined an invariant worldline. If we assume that the spinning particle is in fact spatially extended (as argued by C. M\o ller in an unpublished 1949 manuscript), then a precise position for the particle is a fiction, obtained by averaging over a simultaneity hyperplane. However, the trajectory obtained by averaging over each hyperplane in one foliation of spacetime typically will not agree with the trajectory obtained by averaging over each hyperplane in another foliation. The `failure of covariance' is a symptom of this phenomenon. As Fleming (1965) emphasises, this `failure of covariance' is perfectly well understood, and the hyperplane-dependence can be accommodated so as to define a manifestly Lorentz-covariant Newton-Wigner operator (we need only make explicit the frame for which the position is defined.)

There is an important exception to this hyperplane-dependence, since for massive particles we can demand that the averaging is always over simultaneity hyperplanes defined by the particle's own rest frame. This \emph{does} define an invariant worldline. The resulting position operator Fleming called the `centre of inertia', and showed that it did indeed transform as the spatial part of a Lorentz-covariant four-vector. Indeed, the extra term that shows up in the commutation relations between the Newton-Wigner operator $\mathbf{Q}$ and the Lorentz boost operator $\mathbf{K}$---this extra term is what stops the Newton-Wigner operator from being straightforwardly `covariant', see (\ref{QNextra}) in the appendix---can be understood as a small displacement brought about by a change of foliation into the simultaneity hyperplanes over which the averaging is performed.

However, as Fleming also showed, the components of the `centre of inertia' position operator do not commute among themselves---it is not, in Fleming's terminology, a `local' operator. One might also naturally consider averaging over simultaneity hyperplanes defined by our reference frame---Fleming called this the `centre of mass' position, though it has more recently been dubbed  the `centre of energy' (Schwartz \& Giulini 2020). One finds that this position operator is \emph{neither} `covariant' in the straightforward sense \emph{nor} `local' in Fleming's sense. The Newton-Wigner position operator was dubbed by Fleming the `centre of spin': it is `local' but not `covariant'. Fleming showed that no position operator could be both `covariant' and `local'.

So it seems, for a spinning massive particle, that we do not have a unique position operator at all. It is perhaps reassuring to learn that the three mentioned above agree to within the Compton wavelength of the particle. The contemporary approach to this problem (exemplified by Schwartz \& Giulini 2020) is to decompose the relativistic angular momentum tensor (that is, the generator of rotations and Lorentz boosts) into an orbital and spin component according to
\begin{equation}
M^{\mu\nu} = \xi_f^\mu(\eta, \tau)\cdot P^\nu - \xi_f^\nu(\eta, \tau)\cdot P^\mu + \Sigma_f^{\mu\nu}\ .
\end{equation}
In this decomposition,  $P^\mu$ is the familiar four-momentum. $\xi_f^\mu(\eta, \tau)$ is a candidate position four-vector, for a frame foliated into simultaneity hypersurfaces orthogonal to the constant timelike  vector field $\eta^\mu$ and parameterised by $\tau$ (we thus demand that $\eta_\mu\xi_f^\mu(\eta, \tau) = \tau$).\footnote{Note that, in the proof of our main theorem, we assume such a foliation by presenting the Poincar\'e algebra in the `3+1 split' form, in which we specify $H$, $\mathbf{P}$, $\mathbf{J}$ and $\mathbf{K}$ separately, as opposed to $P^\mu$ and $M^{\mu\nu}$.} $\Sigma_f^{\mu\nu}$ is the spin tensor, constrained by the `supplementary spin condition' (SSC)
\begin{equation}
f_\mu \Sigma_f^{\mu\nu}(\eta) = 0\ ,
\end{equation}
where $f^\mu$ is some timelike vector field (it need not be constant).
This constraint serves also to select the position operator $\xi_f^\mu(\eta, \tau)$. Thus we have infinitely many position operators: and I do not mean due to the frame-relativity determined by $\eta$. Aside from a choice of foliation of spacetime, we have a vast array of position operators \emph{via} the timelike vector field $f^\mu$ that gives us our SSC. Incidentally, the `centre of inertia' corresponds to $f^\mu\propto P^\mu$, the `centre of energy' corresponds to $f^\mu\propto \eta^\mu$, and the Newton-Wigner `centre of spin' corresponds to $f^\mu\propto m\eta^\mu + P^\mu$ (thus we get an apparent failure of covariance if we transform $P^\mu$ while holding $\eta^\mu$ fixed). We implicitly laid down the latter SSC by demanding that the total angular momentum operator $\mathbf{J}$ be decomposable into $\mathbf{L} + \mathbf{S}$, where $\mathbf{S}$ is the particle's intrinsic spin (that is, its spin in its own rest frame); hence `centre of spin'.

Clearly, from the point of view of SSCs, the Newton-Wigner position operator is just one among many possible position operators. But it is still privileged, for it is the unique position operator that has canonical commutation relations as part of the QPS algebra. And in that sense, it is unique in being able to serve as a position operator that is innocent as to dynamics or geometry, since as we have seen the QPS algebra is compatible with a variety of Hamiltonians and geometries (both relativistic and non-relativistic).

\subsection{Field localisation vs.~particle localisation}

We must now turn to the Newton-Wigner position operator's much lamented superluminal propagation problems. There are broadly two views to be found in the literature. The first view, what we might call \emph{field supremacism}, is expressed by Saunders (1994), Malament (1995), Halvorson (2001) and Wallace (2001). According to this view, the superluminal propagation of Newton-Wigner-localised states, and associated failure of microcausality for spacelike-separated regions, torpedos any interpretation according to which Newton-Wigner localisation is the real scheme by which real particles are localised. Insofar as states are really localised, they are localised according to the standards of the underlying quantum field. The second view, what we might call \emph{particle-field ecumenicalism}, upholds that Newton-Wigner localisation is a legitimate localisation scheme---perhaps even \emph{the} legitimate localisation scheme---despite  superluminal propagation, and is defended by Segal (1964), Fleming \& Butterfield (1999) and Fleming (2000).

Common to all parties in this debate is agreement on the fact that the two localisation schemes, Newton-Wigner and field, are not equivalent. Wallace (2001) makes this explicit; I will give just a little more detail here. If we take the 1-particle state (of the free field) $|1_\psi\rangle = a^+(\psi)|0\rangle$, consider the local field operator $\hat{\phi}(\mathbf{x}) = \Phi(\mathbf{0}, -\delta_{\mathbf{x}})$ (see (\ref{fieldop}), above) and using (\ref{FPduality}) with the specification of the 1-particle map
\begin{equation}
K(f, g) = \frac{1}{\sqrt{2\hbar}}\left(\omega^{\frac{1}{2}}f + i\omega^{-\frac{1}{2}}g\right),
\end{equation}
we find that $\hat{\phi}(\mathbf{x}) = \sqrt{\frac{\hbar}{2}}\left(a(\omega^{-\frac{1}{2}}\delta_\mathbf{x}) + a^+(\omega^{-\frac{1}{2}}\delta_\mathbf{x})\right)$, and so
\begin{equation}
\begin{array}{rclrcl}
\langle 0|\hat{\phi}(\mathbf{x})|0\rangle &=& 0\ ;
&
\langle 0|\hat{\phi}(\mathbf{x})^2|0\rangle &=& \frac{1}{2}\hbar\|\omega^{-\frac{1}{2}}\delta_\mathbf{x}\|^2\ ;
\\\\
\langle 1_\psi|\hat{\phi}(\mathbf{x})|1_\psi\rangle &=& 0\ ;
&
\langle 1_\psi|\hat{\phi}(\mathbf{x})^2|1_\psi\rangle &=& \frac{1}{2}\hbar\|\omega^{-\frac{1}{2}}\delta_\mathbf{x}\|^2 + \hbar|\langle \omega^{-\frac{1}{2}}\delta_\mathbf{x}, \psi\rangle|^2
\\
&&&& = &\frac{1}{2}\hbar\|\omega^{-\frac{1}{2}}\delta_\mathbf{x}\|^2  + \hbar|(\omega^{-\frac{1}{2}}\psi)(\mathbf{x})|^2\ ;
\end{array}
\end{equation}
Thus if we choose the 1-particle state $\psi = \delta_\mathbf{y}$, localised at $\mathbf{y}$, the difference between  in expectation values for the square of the local field operator between the field state $|1_\psi\rangle$ and the vacuum is
\begin{equation}
\langle 1_\psi|\hat{\phi}(\mathbf{x})^2|1_\psi\rangle - \langle 0|\hat{\phi}(\mathbf{x})^2|0\rangle = \hbar|(\omega^{-\frac{1}{2}}\delta_\mathbf{y})(\mathbf{x})|^2\ .
\end{equation}
The function on the righthand-side here is only approximately localised at $\mathbf{y}$: it has infinite tails, but is sharply peaked at $\mathbf{y}$ with a width of the order of the Compton wavelength. Thus a particle perfectly localised according the Newton-Wigner scheme corresponds to a merely approximately localised field excitation.

What needs to be explicitly pointed out is that this mismatch between Newton-Wigner localisation and field localisation is precisely what \emph{saves} Newton-Wigner localisation as an \emph{effective} localisation scheme for particles in the eyes of the field supremacist. For it makes explicit why we should not expect local field algebras to contain Newton-Wigner-localised particle occupation numbers, and it gives the precise sense in which such occupation numbers can still be \emph{associated} with a local spatial region $R(t)$, despite not being found in the local algebra $\mathfrak{A}(R(t))$. The mismatch also saves microcausality for the field supremacist; for it makes the small violation of microcausality in the Newton-Wigner localisation scheme compatible with the perfect fulfilment of microcausality for local field operators, and thus answers the challenge posed by Halvorson (2001).

I confess sympathy with field supremacism, and therefore endorse Newton-Wigner localisation as merely an effective localisation scheme for what are, after all, entities in a merely effective ontology. But I must also confess some degree of confusion over the matter. My confusion stems from the following consideration. What counts as the \emph{real} localisation scheme is surely connected to what particle detectors detect. But particle detectors are made of (what else?)~particles, which according to the field picture have states spread over all of space for  all times. We therefore have no good reason to expect ``local'' particle detectors to be perfect measuring devices for observables confined to local field algebras. And I am pessimistic about the matter being resolved experimentally, not just because the violations of microcausality in the Newton-Wigner localisation scheme are so minuscule so as to be effectively invisible for even the most lavishly expensive experiment. A much deeper problem is that it's not  even clear what the field theory  predicts, since any plausible model would have to be specific about the make-up of the local particle detectors, and this will involve making approximations which, it seems to me, would simply lose track of the difference between the two localisation schemes.

\subsection{One field, many particles} \label{1field}

I conclude by owning up to the fact that, while Wigner's identification is surely too narrow, the conception of particles  I have been advocating in this paper is too wide. To be more specific and to take the simplest example: if we take any real scalar field theory, with local field operators $\hat{\phi}(\mathbf{x}), \hat{\pi}(\mathbf{x})$ obeying the usual CCRs, we may analyse it in terms of particles, associated with creation and annihilation operators $a(\psi), a^+(\psi)$, where the 1-particle states $\psi$ live in the unique (thanks for the Stone-von Neumann theorem) irrep of the QP algebra for 3 spatial dimensions (the spin here is zero, so we may ignore $\mathbf{S}$). The choice of particle picture hangs on the choice of the 1-particle map $K$ in (\ref{FPduality}), which takes us from the space of classical field configurations to the 1-particle Hilbert space. The map $K$ is determined by a choice of 1-particle Hamiltonian $\omega$, and this choice is tailor-made for a free field theory in which the field Hamiltonian density is
\begin{equation}\label{Hdens}
\mathcal{H}(\mathbf{x}) = \frac{1}{2}\hat{\pi}(\mathbf{x})^2 + \frac{1}{2}\hat{\phi}(\mathbf{x})\omega^2\hat{\phi}(\mathbf{x})\ .
\end{equation}
This choice of $\omega$ is tailor-made for this free field theory in the sense that the dynamics of the field decomposes perfectly into persisting, non-interacting particles, each evolving according to the 1-particle Hamiltonian $\omega$, and the ground state for the quantum field in this case is given by the Fock vacuum associated with these particles.

But we may still choose the 1-particle map $K$ associated with $\omega$ \emph{even if} the field evolves according to a Hamiltonian density other than (\ref{Hdens}). The associated field ground state will now \emph{not} be the Fock vacuum for these particles, and over time we will find spontaneous production and annihilation of these particles---perhaps over very short timescales. The particles themselves, despite being characterised by the 1-particle Hamiltonian $\omega$, will \emph{not} individually evolve according to this Hamiltonian---indeed, their individual evolution will likely not even be unitary, due to spontaneous production and annihilation. Despite all this, the associated creation and annhilation operators, and consequently also the associated occupation numbers (but typically \emph{not} the total particle number), for these particles will still be well-defined. This would be perfectly analogous to analysing a quantum system according to `bad' quantum numbers: you can do it, but it won't be particularly enlightening. (Incidentally, this is one reason among many for treating the  particle picture as subservient to the field picture. The same quantum field, characterised kinematically by the CCRs, admits of multiple particle pictures. Here I disagree with Baker (2009), since inequivalent representations of the field CCRs can still be conceived of as alternative possibilities for the \emph{same} field, kinematically characterised.)

What is needed, then, is some further restriction on the particle conception, such that the associated choice of particle picture is analogous to a choice of `good' quantum numbers. Wallace (2001) suggests that what we need is some diachronic condition: that the resulting particles last over some reasonably large timescale. Something like this is surely correct, and admits of much further detailed study. My aim in this paper has been merely to argue that the conditions induced by Wigner's identification are far too strict. I hope to have opened up the particle concept so that it may escape the necessity of being non-interacting, and so that we may now concentrate on finding a more suitable narrowing, guided by a consideration of empirically successful, if merely effective, interacting particle theories.

\section{Acknowledgements}

I would like to thank a number of heroes for their wisdom, guidance and encouragement. These include: David Wallace, Bryan Roberts, Owen Maroney, Doreen Fraser, Erik Curiel, Wayne Myrvold, Jeremy Butterfield, Harvey Brown, Simon Saunders, Oliver Pooley, Christopher Timpson, James Read, Valter Moretti, Andre Lukas, Jeffrey Barrett, Ben Feintzeig, Jim Weatherall, Eleanor March and Caspar Jacobs.

\section{Appendix: proof of the main theorem}

Recall that the theorem is as follows:

{\em \noindent Take any massive Wigner representation $(m, s)$ of the Poincar\'e algebra, with generators $H, \mathbf{P}, \mathbf{J}, \mathbf{K}$. Then there are unique $\mathbf{Q}$ and $\mathbf{S}$ such that:
\begin{equation}\label{orbspindec}
\begin{array}{rclcrcl}
 \mathbf{J} &=& \mathbf{L} + \mathbf{S}\ ,& \mbox{where} & \mathbf{L} &:=& \mathbf{Q} \times\mathbf{P}\ ;
 \\
\mathbf{K} &=& \mathbf{M} + \mathbf{N}\ ,& \mbox{where} & \mathbf{M} &:=& t\mathbf{P} - \mathbf{Q}\cdot H \mbox{ and } \mathbf{N.P} = 0\ ; 
\\
\ [\mathbf{Q}, \Lambda] &=& \mathbf{0}\ , & \mbox{where} & \Lambda &:=& H (H^2)^{-\frac{1}{2}}\ .
\end{array}
\end{equation}
and such that either of  the following claims is true:
\begin{enumerate}
\item [(i)] The quantities $H, \mathbf{P}, \mathbf{L}, \mathbf{M}$ obey the Poincar\'e algebra associated with the representation $(m, 0)$.

\item[(ii)]  $\mathbf{Q}$, $\mathbf{P}$ and $\mathbf{S}$ obey the QPS algebra. That is, $\mathbf{Q}$ and $\mathbf{P}$ obey the  Heisenberg algebra; the components of $\mathbf{S}$ obeys the familiar angular momentum algebra;  $\mathbf{Q}$ and $\mathbf{P}$  commute with $\mathbf{S}$; and  $\mathbf{Q}, \mathbf{P}$  and $\mathbf{S}$ act irreducibly on the positive-frequency and negative-frequency subspaces, respectively.
\end{enumerate}
Assuming (\ref{orbspindec}), if either of (i) or (ii) is true, the other one is too.
Moreover,  the following claims are also true:
\begin{itemize}
\item $\mathbf{L}, \mathbf{S}, \mathbf{M}$ and $\mathbf{N}$ are all separately conserved.
\item $\mathbf{Q}$ is the Newton-Wigner operator.
\item $\mathbf{S}$ forms the  unique spin algebra associated with $s$; i.e.~$\mathbf{S}^2 = \hbar^2 s(s+1)$.
\item  The generators $H, \mathbf{P}, \mathbf{J}, \mathbf{K}$ take the Foldy form; in particular $\mathbf{N}$ is of the form $\mathbf{N} = \frac{\Lambda\mathbf{S}\times\mathbf{P}}{\omega + m}$.
\item $\mathbf{Q}$ transforms covariantly under $\mathbf{M}$ (but  not $\mathbf{K}$ if $\mathbf{S} \neq\mathbf{0}$).
\end{itemize}
}

The existence claim is relatively straightforward to prove: we simply take $\mathbf{Q}$ to be the Newton-Wigner operator, and appeal to the Foldy (1956) form for the generators $H, \mathbf{P}, \mathbf{J}, \mathbf{K}$. Foldy proved that, given this form for the Poincar\'e generators, with the usual QPS algebraic relations, the Poincar\'e relations are realised.

\begin{proposition}\label{Lemma-N!}
If the orbital angular momentum is conserved, $[\mathbf{L}, H] = \mathbf{0}$, then the internal boost operator $\mathbf{N}$ can be given  the precise form $\displaystyle \mathbf{N} = \frac{\Lambda\mathbf{S}\times\mathbf{P}}{\omega + m}$.
\end{proposition}

\noindent\emph{Proof.} 
This was stated above in the previous section, but the details are as follows. From (\ref{PLspin}) above we have that the spatial part of the Pauli-Lubanski pseudovector is
\begin{equation}\label{WS}
\mathbf{W} = H\mathbf{S} - \mathbf{P}\times\frac{\mathbf{S}\times\mathbf{P}}{H+m}
\end{equation}
Plugging (\ref{orbspindec}) into the standard definition of $\mathbf{W}$ as $\mathbf{W} = H\mathbf{J} - \mathbf{P}\times\mathbf{K}$, we find that
\begin{equation}
\mathbf{W} = -[\mathbf{L}, H] + H\mathbf{S} - \mathbf{P}\times\mathbf{N}\ .
\end{equation}
Thus if $[\mathbf{L}, H] = \mathbf{0}$, then $\mathbf{P}\times\mathbf{N} = \mathbf{P}\times\frac{\mathbf{S}\times\mathbf{P}}{H+m}$. Since we additionally assume that $\mathbf{N.P} = 0$, it follows that $\mathbf{N} = \frac{\Lambda\mathbf{S}\times\mathbf{P}}{\omega +m}$. $\Box$

\begin{proposition}\label{Lemma1}
 (\ref{orbspindec}) entails that (i) and (ii)  are equivalent.
\end{proposition}
The (ii) $\Rightarrow$ (i) direction is once again given by Foldy (1956) for the spinless case. The (i) $\Rightarrow$ (ii) direction is the most substantial proposition (in the sense that I believe no other proof has yet been given), and will be proved using a collection of lemmas. So, to clarify: our plan here is to assume that $H, \mathbf{P}, \mathbf{J}, \mathbf{K}$ generate the  Wigner representation $(m, s)$, where $m > 0$, and also that $H, \mathbf{P}, \mathbf{L}, \mathbf{M}$ generate the spinless Wigner representation $(m, 0)$ with the same mass, and from this to derive the canonical QPS algebraic relations.

The chain of lemmas begins with a consideration of the velocity operator $\mathbf{V} := \frac{1}{i\hbar}[\mathbf{Q}, H]$.

\begin{lemma}\label{Lemma-velpar}
The velocity is parallel to the momentum; i.e.~$\mathbf{V} = \kappa\mathbf{P}$ for some undetermined factor $\kappa$.
\end{lemma}

\noindent \emph{Proof.}
We consider the commutation relations between $\mathbf{L}$ and $H$:
\begin{equation}
\ [L_i, H] = \varepsilon_{ijk}[Q_j P_k, H]
= \varepsilon_{ijk}[Q_j, H] P_k
= i\hbar\varepsilon_{ijk}V_jP_k\ .
\end{equation}
We have applied the conservation of momentum, as dictated by the Poincar\'e relations between $\mathbf{P}$ and $H$. The Poincar\'e relations for $\mathbf{L}$ and $H$ dictate that $\mathbf{L}$ is conserved; i.e.~$[\mathbf{L}, H] = 0$. It immediately follows that $V_1P_2 = V_2P_1, V_1P_3 = V_3P_1$ and $V_2P_3 = V_3P_2$. Since this holds for any value of $\mathbf{P}$, we must have $\mathbf{V} = \kappa \mathbf{P}$, for some undetermined self-adjoint factor $\kappa$.
 $\Box$

\begin{lemma}\label{Lemma-XP}
The position and momentum obey the standard Heisenberg relations
$[Q_i, P_j] = i\hbar\delta_{ij}$ and the velocity is translation-invariant: $[{V}_i, P_j] = 0$.
\end{lemma}

\noindent\emph{Proof.}
We consider the commutation relations between $\mathbf{L}$ and $\mathbf{P}$.
\begin{equation}
\ [L_i, P_j] = \varepsilon_{ikl}[Q_k P_l, P_j]
= \varepsilon_{ikl}[Q_k, P_j] P_l\ .
\end{equation}
We have used the fact that the components of $\mathbf{P}$  commute among themselves. According to the Poincar\'e relations, $\mathbf{P}$ transforms under rotations as a 3-vector; i.e.~ 
$ [L_i, P_j] = i\hbar\varepsilon_{ijk}P_k$.
Combining these equations we obtain
\begin{equation}\label{Heisenberg0}
\begin{array}{ccc}
\ [Q_2, P_1]P_3 - [Q_3, P_1]P_2 = 0\ ;
& \qquad &
\ [Q_2, P_2]P_3 - [Q_3, P_2]P_1 = i\hbar P_3\ ;
\\
\ [Q_3, P_2]P_1 - [Q_1, P_2]P_3 = 0\ ;
& \qquad &
\ [Q_3, P_3]P_1 - [Q_1, P_3]P_2 = i\hbar P_1\ ;
\\
\ [Q_1, P_3]P_2 - [Q_2, P_3]P_1 = 0\ ;
& \qquad &
\ [Q_1, P_1]P_2 - [Q_2, P_1]P_3 = i\hbar P_2\ .
\end{array}
\end{equation}
These identities hold for arbitrary $\mathbf{P}$. So the commutation relations must be of the form
\begin{equation}
\begin{array}{rclrclrcl}
\ [Q_1, P_1] &=& i\hbar + \xi_1 P_3\ ;
&
\ [Q_1, P_2] &=& \xi_2 P_1\ ;
&
\ [Q_1, P_3] &=& \xi_3 P_1\ ;
\\
\ [Q_2, P_1] &=& \xi_1 P_2\ ;
&
\ [Q_2, P_2] &=& i\hbar + \xi_2 P_1\ ;
&
\ [Q_2, P_3] &=& \xi_3 P_2\ ;
\\
\ [Q_3, P_1] &=& \xi_1 P_3\ ;
&
\ [Q_3, P_2] &=& \xi_2 P_3\ ;
&
\ [Q_3, P_3] &=& i\hbar + \xi_3 P_2 \ ;
\end{array}
\end{equation}
where $\xi_1, \xi_2, \xi_3$ are factors to be determined.

Now consider that $\mathbf{Q}, \mathbf{P}$ and $\mathbf{L}$ are all self-adjoint. Then
\begin{equation}
L_i^\dag = \left(\varepsilon_{ijk}Q_jP_k\right)^\dag  = \varepsilon_{ijk}P^\dag_k Q^\dag_j
= \varepsilon_{ijk}P_k Q_j =  \varepsilon_{ijk}Q_j P_k  = L_i\ .
\end{equation}
The crucial identity is  the fourth one. For the $L_1$ component we have:
\begin{equation}
Q_2P_3 - Q_3P_2 = P_3Q_2 - P_2Q_3
\qquad\Rightarrow\qquad
[Q_2, P_3] = [Q_3, P_2]\ ,
\end{equation}
which entails that $\xi_2P_3 = \xi_3P_2$. We similarly obtain $\xi_3P_1 = \xi_1P_3$ and $\xi_1P_2 = \xi_2P_1$. It follows that $\bm{\xi} = i\hbar\eta \mathbf{P}$ for some undetermined self-adjoint factor $\eta$ (the appearance of $\hbar$ is mere convenience). And so we have the identities
\begin{equation}
\begin{array}{rclrclrcl}
\ [Q_1, P_1] &=& i\hbar(1 + \eta P_1 P_3)\ ;
&
\ [Q_1, P_2] &=& i\hbar\eta P_1P_2\ ;
&
\ [Q_1, P_3] &=& i\hbar\eta P_1P_3\ ;
\\
\ [Q_2, P_1] &=& i\hbar\eta P_1 P_2\ ;
&
\ [Q_2, P_2] &=& i\hbar(1 + \eta P_1P_2)\ ;
&
\ [Q_2, P_3] &=& i\hbar\eta P_2P_3\ ;
\\
\ [Q_3, P_1] &=& i\hbar\eta P_1P_3\ ;
&
\ [Q_3, P_2] &=& i\hbar\eta P_2 P_3\ ;
&
\ [Q_3, P_3] &=& i\hbar(1 + \eta P_2P_3) \ .
\end{array}
\end{equation}

Now consider the commutation relations between the $[Q_i, P_j]$ and $H$. We obtain
\begin{equation}
\left[[Q_i, P_j], H\right] = i\hbar[V_i, P_j ] = i\hbar[\kappa, P_j]P_i\ .
\end{equation}
We have used Lemma \ref{Lemma-velpar} in the last step.
This generates the six identities (we expect nine identities, but three are redundant):
\begin{equation}
\begin{array}{rclrclrcl}
\ [\eta, H] P_3 &=& [\kappa, P_1]\ ;
&
\ [\eta, H] P_2 &=& [\kappa, P_2]\ ;
&
\ [\eta,H] P_3 &=& [\kappa, P_3]\ ;
\\
\ [\eta,H] P_1 &=& [\kappa, P_1]\ ;
&
\ [\eta,H] P_1 &=& [\kappa, P_2]\ ;
&
\ [\eta, H] P_2 &=& [\kappa, P_3]\ .
\end{array}
\end{equation}
It immediately follows that $[\eta, H] = 0$ and $[\kappa, \mathbf{P}] = \mathbf{0}$.  It follows from the latter identity, with Lemma \ref{Lemma-velpar}, that the velocity $\mathbf{V}$ is translation invariant; i.e.~$[V_i, P_j] = 0$.

We can further pin down the factor $\eta$. Consider the commutation relations between $\mathbf{M}$ and $\mathbf{P}$:
\begin{eqnarray}
[M_i, P_j] &=& -\frac{1}{c^2}\left[Q_i\cdot H, P_j\right]
\nonumber\\
&=& -\frac{1}{c^2} [Q_i, P_j]\cdot H\ .
\end{eqnarray}
We have used the conservation of momentum, $[\mathbf{P}, H] = 0$.
According to the Poincar\'e relations, $[M_i, P_j] = -i\hbar\frac{1}{c^2}\delta_{ij}H$; so for any  $i \neq j$ we have, using $[\mathbf{P}, H] = \mathbf{0}$ and $[\eta, H] = 0$,
\begin{equation}
\{\eta, H\}P_iP_j = 0\ ,
\end{equation}
and we can see that  $\eta$ anticommutes with the Hamiltonian: $
\{\eta, H\} = 0$.
But if $\eta$ both commutes and anti-commutes with $H$, then $\eta H = 0$, and so $\eta = 0$.  We thereby finally recover the standard Heisenberg relations between position and momentum, $[Q_i, P_j] = i\hbar\delta_{ij}$. $\Box$

\begin{lemma}\label{Lemma-VH}
The velocity $\mathbf{V}$ is conserved; i.e.~$[\mathbf{V}, H] = \mathbf{0}$\ .
\end{lemma}

\noindent\emph{Proof.}
We first use Lemma \ref{Lemma-XP} and the fact that $\omega$ is a function of $\mathbf{P}$. Then $[\mathbf{V}, \omega] = 0$, and so
\begin{equation}
\ [V_i, H] = [V_i, \Lambda\omega]
= [V_i, \Lambda]\omega \ .
\end{equation}
Now recall that $\mathbf{V} = \frac{1}{i\hbar}[\mathbf{Q}, H]$ by definition. We use the fact that $[\Lambda, H] = 0$ and $[\mathbf{Q}, \Lambda] = \mathbf{0}$. Then
\begin{equation}
\ [V_i, \Lambda] = \frac{1}{i\hbar}\left[[Q_i, H], \Lambda\right]
= \frac{1}{i\hbar}\left[[Q_i, \Lambda], H\right]
= 0\ .
\end{equation}
It then follows that $[\mathbf{V}, H] = \mathbf{0}$. $\Box$

\begin{lemma}\label{Lemma-V}
The velocity takes the standard form $\mathbf{V} = c^2\mathbf{P}H^{-1}$.
\end{lemma}

\noindent\emph{Proof.} 
We consider the commutation relations between $\mathbf{M}$ and $H$:
\begin{equation}
\ [M_i, H] = -\frac{1}{2c^2}\left([Q_i, H]H + H[Q_i, H]\right)
= -i\hbar\frac{1}{c^2}V_i\cdot H
\ .
\end{equation}
In the first identity we have applied the conservation of momentum, as dictated by the Poincar\'e relations between $\mathbf{P}$ and $H$, and
in the last identity we have applied the definition of $\mathbf{V}$. Now applying Lemma \ref{Lemma-VH} we obtain $[M_i, H] = -i\hbar\frac{1}{c^2}V_i H$.  The Poincar\'e relations demand that $[M_i, H] = -i\hbar P_i$; it follows that $\mathbf{V} = c^2\mathbf{P}H^{-1}$. $\Box$

\begin{lemma}\label{Lemma-XX}
The components of the position operator $\mathbf{x}$ commute among themselves: $[Q_i, Q_j] = 0$.
\end{lemma}

\noindent\emph{Proof.}
We consider the commutation relations between the components of $\mathbf{M}$. We use the following form for $\mathbf{M}$:
\begin{equation}
\mathbf{M} \ =\  t\mathbf{P} - \frac{1}{c^2}\mathbf{Q}H + \frac{1}{2c^2}[\mathbf{Q}, H]
\ =\ t\mathbf{P} - \frac{1}{c^2}\mathbf{Q}H + i\hbar\frac{1}{2c^2}\mathbf{V}
\ =\ t\mathbf{P} - \frac{1}{c^2}\mathbf{Q}H + \frac{1}{2}i\hbar\mathbf{P}H^{-1}\ ,
\end{equation}
where in the last identity we have applied Lemma \ref{Lemma-V}. Then, assuming the Poincar\'e relations for $\mathbf{P}$ and $H$, we obtain
\begin{eqnarray}
[M_i, M_j] &=& \frac{1}{c^4}\left([Q_i, Q_j]H^2 + Q_j[Q_i, H]H - Q_i[Q_j, H]H\right)  - \frac{1}{c^2}t\left([Q_i, P_j] - [Q_j, P_i]\right)H
\nonumber\\
&& \qquad -\ i\hbar\frac{1}{2c^2}\left([Q_i, P_j] - [Q_j, P_i] + P_j[Q_i, H^{-1}]H - P_i[Q_j, H^{-1}]H\right)\ .
\end{eqnarray}
We now apply the identity
\begin{equation}
[\mathbf{Q}, H^{-1}] = -H^{-1}[\mathbf{Q}, H]H^{-1} = -i\hbar H^{-1}\mathbf{V}H^{-1}
\end{equation}
and  Lemmas \ref{Lemma-XP} and \ref{Lemma-V}. The only surviving terms are
\begin{equation}
[M_i, M_j]\ =\ \frac{1}{c^4}[Q_i, Q_j]H^2 - i\hbar\frac{1}{c^2}\left(Q_iP_j - Q_jP_i\right)
\ = \
\frac{1}{c^4}[Q_i, Q_j]H^2 - i\hbar\frac{1}{c^2}\varepsilon_{ijk}L_k\ .
\end{equation}
The Poincar\'e relations demand that $[M_i, M_j] = -i\hbar\frac{1}{c^2}\varepsilon_{ijk}L_k$. It follows that  $[Q_i, Q_j]H^2 = 0$ and so $[Q_i, Q_j] = 0$. $\Box$

\begin{lemma}\label{Lemma-XL}
The position operator $\mathbf{Q}$ behaves as a vector under rotations generated by $\mathbf{L}$; i.e.~$[Q_i, L_j] = i\hbar\varepsilon_{ijk}Q_k$.
\end{lemma}

\noindent{Proof.} This follows straightforwardly from the Heisenberg relations $[Q_i, P_j] = i\hbar\delta_{ij}$ and $[Q_i, Q_j] = [P_i, P_j] = 0$:
\begin{eqnarray}
[Q_i, L_j] &=& \varepsilon_{jkl}[Q_i, Q_kP_l]
\nonumber\\
&=& \varepsilon_{jkl}\left([Q_i, Q_k]P_l + Q_k[Q_i, P_l]\right)
\nonumber\\
&=& i\hbar\varepsilon_{jkl}\delta_{il} Q_k
\nonumber\\
&=& i\hbar\varepsilon_{ijk} Q_k\ . \qquad \Box
\end{eqnarray}

Incidentally, this result is sufficient to derive  the Poincar\'e relations between the components of $\mathbf{L}$ and between $\mathbf{L}$ and $\mathbf{M}$ without having to assume them, since these relations essentially say that $\mathbf{L}$ and $\mathbf{M}$ act as vectors under rotations generated by $\mathbf{L}$.

\begin{lemma}\label{Lemma-SP}
The momentum and spin operators commute: $[S_i, P_j] = 0$.
\end{lemma}

\noindent\emph{Proof.}
We  compare the commutation relations $[L_i, P_j]$ and $[J_i, P_j]$.
\begin{equation}
\ [J_i, P_j] = [L_i, P_j] + [S_i, P_j] = i\hbar\varepsilon_{ijk}P_k
\end{equation}
Given the Poincar\'e relations between $\mathbf{L}$ and $\mathbf{P}$, the result immediately follows.
$\Box$

\begin{lemma}\label{Lemma-SH}
The spin $\mathbf{S}$ is  conserved; i.e.~$[\mathbf{S}, H] = \mathbf{0}$.
\end{lemma}

\noindent\emph{Proof.}
This follows from the conservation of $\mathbf{J} = \mathbf{L} + \mathbf{S}$ and of $\mathbf{L}$ alone, both of which follow from the Poincar\'e relations. $\Box$

\begin{corollary}\label{Lemma-SLambda}
The spin $\mathbf{S}$ is even; i.e.~$[\mathbf{S}, \Lambda] = \mathbf{0}$.
\end{corollary}

\begin{lemma}\label{Lemma-NP}
The internal boost operator $\mathbf{N}$ is translation invariant; i.e.~$[N_i, P_j] = 0$.
\end{lemma}

\noindent\emph{Proof.}
We  compare the commutation relations $[M_i, P_j]$ and $[K_i, P_j]$:
\begin{equation}
\ [K_i, P_j] = [M_i, P_j] + [N_i, P_j] = -i\hbar\frac{1}{c^2}\delta_{ij}H
\end{equation}
Given the Poincar\'e relations between $\mathbf{M}$ and $\mathbf{P}$, the result immediately follows.
$\Box$

\begin{lemma}\label{Lemma-MNH}
$\mathbf{M}$ and $\mathbf{N}$ are separately conserved: i.e.~$\frac{\textrm{\scriptsize d}\mathbf{M}}{\textrm{\scriptsize d}t} = \mathbf{0}$ and $[\mathbf{N}, H] = \mathbf{0}$.
\end{lemma}

\noindent\emph{Proof.}
Using the conservation of momentum, $[\mathbf{P}, H] = 0$ we obtain
\begin{equation}
[M_i, H] = -\frac{1}{c^2}\left[Q_i, H\right]H = -i\hbar P_i\ .
\end{equation}
Due to $\mathbf{M}$'s explicit time dependence, we have $\frac{\partial \mathbf{M}}{\partial t} = \mathbf{P}$. It follows that
\begin{equation}
\frac{\textrm{d}\mathbf{M}}{\textrm{d}t}\ =\ \frac{\partial \mathbf{M}}{\partial t} + \frac{1}{i\hbar}[M_i, H]\ =\ \mathbf{0},
\end{equation}
and so $\mathbf{M}$ is conserved.
This and the Poincar\'e relation $[\mathbf{K}, H] = -i\hbar\mathbf{P}$ then entails that $[\mathbf{N}, H] = \mathbf{0}$.
$\Box$

\begin{corollary}\label{Lemma-NLambda}
The intrinsic boost operator $\mathbf{N}$ is even; i.e.~$[\mathbf{N}, \Lambda] = \mathbf{0}$.
\end{corollary}

\noindent\emph{Proof.}
From Lemmas \ref{Lemma-NP} and \ref{Lemma-MNH}, $\mathbf{N}$ commutes with both $\mathbf{P}$ and ${H}$. But $H = \Lambda\omega$ and $\omega$ is a function only of $\mathbf{P}$. It follows that $[\mathbf{N}, \omega] =0$, and so $[\mathbf{N}, \Lambda] = 0$. $\Box$

\begin{lemma}\label{Lemma-SJ}
The intrinsic spin $\mathbf{S}$ behaves as a vector under rotations generated by $\mathbf{J}$; i.e.~$[S_i, J_j] = i\varepsilon_{ijk}S_k$.
\end{lemma}

\noindent\emph{Proof.} We use the fact that $\mathbf{S}$ is the following linear combination of the spatial part of the Pauli-Lubanski pseudovector $\mathbf{W}$ and the momentum $\mathbf{P}$:
\begin{equation}
\mathbf{S} = \frac{\mathbf{W}}{m} - \frac{W_0\mathbf{P}}{m(\omega+m)}\ ,
\end{equation}
which follows from the definition of $W_\mu$ and (\ref{WS}). (You can think of $\mathbf{S}$ as obtained from $\frac{W_\mu}{m}$ by Lorentz-transforming from the lab frame into the rest frame of the particle.) The quantities $W_0$, $m$ and $\omega$ all commute with $\mathbf{J}$, and $\mathbf{W}$ and $\mathbf{P}$ behave as vectors under rotations generated by $\mathbf{J}$; therefore $\mathbf{S}$ does too. $\Box$

\begin{corollary}\label{Lemma-NJ}
The intrinsic boost operator $\mathbf{N}$ acts as a vector under rotations generated by $\mathbf{J}$; i.e.~$[N_i, J_j] = i\hbar\varepsilon_{ijk}N_k$.
\end{corollary}

\noindent\emph{Proof.}
This follows straightforwardly from the explicit form of $\mathbf{N}$ given by Proposition \ref{Lemma-N!}, Lemma \ref{Lemma-SJ}, the fact that  $m$ and $\omega$  both commute with $\mathbf{J}$ and the fact that $\mathbf{P}$ behaves as a vector under rotations generated by $\mathbf{J}$.
$\Box$

\begin{lemma}\label{Lemma-SX}
The position and spin operators commute; i.e.~$[Q_i, S_j] = 0$.
\end{lemma}

\noindent\emph{Proof.}
We  compare the commutation relations $[K_i, J_j]$ and $[M_i, L_j]$:
\begin{eqnarray}
[K_i, J_j]&=& [M_i, L_j] + [M_i, S_j] + [N_i, J_j]
\nonumber\\
&=& i\hbar\varepsilon_{ijk}\left(M_k + N_k\right)  + [M_i, S_j]
\nonumber\\
&= & i\hbar\varepsilon_{ijk}K_k + [M_i, S_j]\ .
\end{eqnarray}
(We have implemented Corollary \ref{Lemma-NJ}.)
The Poincar\'e relations between $\mathbf{K}$ and $\mathbf{J}$ dictate that $[K_i, J_j] =  i\hbar\varepsilon_{ijk}K_k$. It follows that $[M_i, S_j] = 0$.  Using the translation invariance and conservation of $\mathbf{S}$, we obtain the condition
\begin{equation}
\ [M_i, S_j] = -\frac{1}{c^2}[Q_iH, S_j] = -\frac{1}{c^2}[Q_i, S_j] H = 0\ .
\end{equation}
But $H$ is never zero. It follows that $[Q_i, S_j] = 0$.
$\Box$

\begin{corollary}\label{Lemma-SL}
The spin  $\mathbf{S}$ commutes with the orbital angular momentum $\mathbf{L}$; i.e.~$[S_i, L_j] = 0$.
\end{corollary}

\noindent\emph{Proof.}
This is immediate from the form of $\mathbf{L}$ in terms of $\mathbf{Q}$ and $\mathbf{P}$ and Lemmas \ref{Lemma-SP} and \ref{Lemma-SX}. $\Box$

\begin{lemma}\label{Lemma-SS}
The components of the spin  $\mathbf{S}$ obey the canonical spin algebra; i.e.~$[S_i, S_j] = i\hbar\varepsilon_{ijk}S_k$, and $\mathbf{S}^2 = \hbar^2 s(s+1)$.
\end{lemma}

\noindent\emph{Proof.}
For this we consider the Poincar\'e relations between the components of the full angular momentum $\mathbf{J}$ and compare them to those of the orbital angular momentum $\mathbf{L}$. We have that
\begin{eqnarray}
[J_i, J_j] &=& [L_i, L_j] + [L_i, S_j] + [S_i, L_j] + [S_i, S_j]
\nonumber
\\
&=& [L_i, L_j] +  [S_i, S_j]
\nonumber
\\
\Rightarrow \qquad
i\hbar\varepsilon_{ijk}J_k &=&i\hbar\varepsilon_{ijk}L_k  +  [S_i, S_j]
\nonumber
\\
\Rightarrow \qquad
[S_i, S_j]  &=& i\hbar\varepsilon_{ijk}S_k\ .
\end{eqnarray}
(We use Corollary \ref{Lemma-SL} in the second identity.) Since we assume that $H, \mathbf{P}, \mathbf{J}, \mathbf{K}$ generate the Wigner representation $(m, s)$, it follows that $W_\mu W^\mu = -\hbar^{2}m^2s(s+1)$. But given the form of the decomposition (\ref{orbspindec}), we have $W_\mu W^\mu = -m^{2}\mathbf{S}^2$; it follows that  $\mathbf{S}^2 = \hbar^2 s(s+1)$.
$\Box$

\begin{proposition}\label{Lemma-XPS}
$\mathbf{Q}, \mathbf{P}$ and $\mathbf{S}$ act irreducibly on the positive-frequency subspace and on the negative-frequency subspace, respectively.
\end{proposition}

\noindent\emph{Proof sketch.}
Start with the fact that, by assumption, $H, \mathbf{P}, \mathbf{J}, \mathbf{K}$ act irreducibly on each of the positive- and negative-frequency Wigner representations associated with $(m, s)$. Then show that any even operator $E$ which commutes with all of $\mathbf{Q}, \mathbf{P}$ and $\mathbf{S}$ must commute with all of  $H, \mathbf{P}, \mathbf{J}, \mathbf{K}$; it follows that $E$ is a multiple of the identity, and so $\mathbf{Q}, \mathbf{P}$ and $\mathbf{S}$ is irreducible on the positive- and negative-frequency subspaces. This is tedious but straightforward, given the decomposition (\ref{orbspindec}).
$\Box$

This concludes our derivation of the QPS algebraic relations. We now turn to Lorentz-transformation properties of the position operator $\mathbf{Q}$ and the fact that $\mathbf{Q}$ must be the Newton-Wigner operator.

\begin{proposition}\label{Lemma2}
$\mathbf{Q}$ transforms like the spatial part of a Lorentz four-vector under $\mathbf{M}$, i.e.~$[Q_i, M_j] = i\hbar(\delta_{ij}t - \frac{1}{c^2}Q_j\cdot V_i)$, where $\mathbf{V}$ is the velocity operator, and transforms similarly under $\mathbf{K}$  only if $s = 0$ (in which case $\mathbf{K} = \mathbf{M}$).
\end{proposition}

\noindent\emph{Proof.} First, it might help to explain why we expect the commutation relations $[Q_i, M_j] = i\hbar(\delta_{ij}t - \frac{1}{c^2}Q_j\cdot V_i)$. For a small Lorentz transformation by velocity $\mathbf{u}$, we have, to first order in $\frac{\mathbf{u}}{c}$,
\begin{equation}
\mathbf{Q}'(t') = \mathbf{Q}(t) + \mathbf{u}t\ ; 
\qquad
t' = t + \frac{\mathbf{u}.\mathbf{Q}(t)}{c^2}\ .
\end{equation}
What we want to know is the form of $\mathbf{Q}'(t)$, for the original time $t$. So we invert the second identity for $t'$ and plug into the first identity to obtain, again to first order,
\begin{eqnarray}
\mathbf{Q}'(t) &=&\mathbf{Q}'\left(t' - \frac{\mathbf{u}.\mathbf{Q}(t)}{c^2}\right)
\nonumber
\\
&=&\mathbf{Q}'(t') - \frac{\mathbf{u}.\mathbf{Q}(t)}{c^2}\cdot\dot{\mathbf{Q}}'(t')
\nonumber
\\
&=&\mathbf{Q}(t) + \mathbf{u}t - \frac{1}{c^2}(\mathbf{u}.\mathbf{Q}(t))\cdot\mathbf{V}(t)\ .
\end{eqnarray}
Thus
\begin{equation}
\frac{\partial Q_i}{\partial u_j} = \delta_{ij}t - \frac{1}{c^2}Q_j\cdot V_i\ .
\end{equation}

Now we turn to establishing the correct commutation relations. This is a simple matter of decomposing $\mathbf{M}$ according to (\ref{orbspindec}) and applying the QPS algebraic relations:
\begin{eqnarray}
[Q_i, M_j]  &=& [Q_i, P_j]t - \frac{1}{c^2}[Q_i, Q_j\cdot H]
\nonumber
\\
&=& [Q_i, P_j]t -  \frac{1}{c^2}Q_j\cdot [Q_i, H]
\nonumber
\\
&=& i\hbar\left(\delta_{ij}t -  \frac{1}{c^2}Q_j\cdot V_i\right)\ ,
\end{eqnarray}
which is the first claim we wished to prove. 

We now look at the transformation of $\mathbf{Q}$ under the ``full'' boost operator $\mathbf{K}$. We find that
\begin{equation}
[Q_i, K_j] = [Q_i, M_j] + [Q_i, N_j] \ .
\end{equation}
Now, given the form of $\mathbf{N}$ given in Proposition \ref{Lemma-N!}, we find that
\begin{equation}
[Q_i, N_j] =
\varepsilon_{jkl}\frac{S_k}{\omega+m}[Q_i, P_l] + \varepsilon_{jkl}S_kP_l\left[Q_i, \frac{1}{\omega + m}\right]
\end{equation}
We now use the fact that $[A, B^{-1}] = -B^{-1}[A, B]B^{-1}$ and $[\mathbf{Q}, \omega] = i\hbar\mathbf{V} = i\hbar \mathbf{P}\omega^{-1}$ to obtain
\begin{equation}\label{QNextra}
[Q_i, N_j] =
i\hbar\left(\frac{\varepsilon_{ijk}S_k}{\omega+m} - \frac{P_i N_j}{\omega(\omega + m)}\right)
\end{equation}
It should then be clear that $[Q_i, N_j] \neq 0$ unless $\mathbf{S} = \mathbf{0}$. Thus  $\mathbf{Q}$ transforms like the spatial part of a Lorentz four-vector under $\mathbf{K}$ if and only if $\mathbf{S} = \mathbf{0}$.
$\Box$

\begin{proposition}\label{Lemma2}
The Newton-Wigner operator $\mathbf{Q}$ satisfies (\ref{orbspindec}), (i) and (ii) above.
\end{proposition}

\noindent\emph{Proof sketch.}
It suffices to show that $\mathbf{Q}$ satisfies  (\ref{orbspindec}) and (ii); we  then apply Proposition \ref{Lemma1}. That $\mathbf{Q}$ satisfies  (\ref{orbspindec}) can be easily seen from the Foldy form (1956); we then need only to appeal to the well-known features of $\mathbf{Q}$, which include (ii). $\Box$

\begin{proposition}\label{Lemma3}
 Only the Newton-Wigner operator $\mathbf{Q}$ satisfies both (\ref{orbspindec}) and (ii)  above.
\end{proposition}

\noindent\emph{Proof sketch.}
This was proved by Jordan (1980). The general idea is to define  $\mathbf{R} = \mathbf{Q} + \bm{\Delta}$, assume the same commutation relations on $\mathbf{R}$ as on $\mathbf{Q}$, to then establish that $\bm{\Delta} = \mathbf{0}$. $\Box$

\begin{proposition}\label{Lemma4}
 Only the Newton-Wigner operator $\mathbf{Q}$ satisfies both (\ref{orbspindec}) and (i)  above.
\end{proposition}

\noindent\emph{Proof sketch.}
This follows from  Propositions \ref{Lemma1} and \ref{Lemma3}, but was essentially proved by Jordan \& Mukunda (1963). $\Box$

\section{References}

\begin{itemize}
\item Baker, D. J. (2009). Against Field Interpretations of Quantum Field Theory. The British Journal for the Philosophy of Science, 60, pp.~585-609.
\item Bain, J. (2000). Against particle/field duality: Asymptotic particle states and interpolating fields in interacting QFT (or: Who's afraid of Haag's theorem?). Erkenntnis, 53(3), 375-406.
\item Baez, J. C., Segal, I. E., \& Zhou, Z. (1992). Introduction to algebraic and constructive quantum field theory. Princeton University Press.
\item Castellani, E. (2002). Symmetry, quantum mechanics, and beyond. Foundations of Science, 7, 181-196.
\item Crater, H. W., \& Van Alstine, P. (1999). Two-body Dirac equations for relativistic bound states of quantum field theory. arXiv preprint hep-ph/9912386.
\item Duncan, A. (2012). The conceptual framework of quantum field theory. Oxford University Press.
\item Earman, J. (1989). World enough and spacetime.
\item Earman, J., \& Fraser, D. (2006). Haag?s theorem and its implications for the foundations of quantum field theory. Erkenntnis, 64(3), 305-344.
\item Foldy, L. L. (1956). Synthesis of covariant particle equations. Physical Review 102.2 (1956): 568.
\item Fleming, G. N. (1965). Covariant position operators, spin, and locality. Physical Review, 137(1B), B188.
\item Fleming, G. N. (2000). Reeh-Schlieder meets Newton-Wigner. Philosophy of Sci- ence 67 (Proceedings), S495?S515.
\item Fleming, G. N. \& Butterfield, J. (1999). Strange positions. From physics to philosophy, 108-165.
\item Fraser, D. (2008). The fate of ?particles? in quantum field theories with interactions. Studies in History and Philosophy of Science Part B: Studies in History and Philosophy of Modern Physics, 39(4), 841-859.
\item Gell-Mann, M., \& Low, F. (1951). Bound states in quantum field theory. Physical Review, 84(2), 350.
\item Gorelick, J. L., \& Grotch, H. (1977). Derivation of an effective one-body Dirac equation from the Bethe-Salpeter equation. Journal of Physics G: Nuclear Physics, 3(6), 751.
\item Halvorson, H. (2001). Reeh-Schlieder defeats Newton-Wigner: On alternative localization schemes in relativistic quantum field theory. Philosophy of Science, 68(1), 111-133.
\item Hegerfeldt, G. C. (1974). Remark on causality and particle localization. Physical Review D, 10(10), 3320.
\item Jordan, T. F. (1980). Simple derivation of the Newton?Wigner position operator. Journal of Mathematical Physics, 21(8), 2028-2032.
\item Jordan, T. F., \& Mukunda, N. (1963). Lorentz-covariant position operators for spinning particles. Physical Review, 132(4), 1842.
\item Lindgren, I., Salomonson, S., \& Hedendahl, D. (2005). Many-body-QED perturbation theory: Connection to the two-electron Bethe Salpeter equation. Canadian journal of physics, 83(3), 183-218.
\item Levy-Leblond, J. M. (1963). Galilei group and nonrelativistic quantum mechanics. Journal of mathematical Physics, 4(6), 776-788.
\item Mackey, G. W. (1951). On induced representations of groups. American Journal of Mathematics, 73(3), 576-592.
\item Malament, David B. (1995), `In defense of dogma: Why there cannot be a relativistic quantum mechanics of (localizable) particles.' UNIVERSITY OF WESTERN ONTARIO SERIES IN PHILOSOPHY OF SCIENCE: 1-10.
\item Ne'eman, Y., Sternberg, S., Donato, P., Duval, C., Elhadad, J., \& Tuynman, G. (1991). Internal supersymmetry and superconnections. Symplectic Geometry and Mathematical Physics, 326-54.
\item Newton, T. D., \& Wigner, E. P. (1949). Localized states for elementary systems. Reviews of Modern Physics, 21(3), 400.
\item Redhead, M. L. G. (1995), `More Ado About Nothing', Foundations of Physics 25: 123-137.
\item Saunders, S. (1994). A dissolution of the problem of locality. In PSA: Proceedings of the Biennial Meeting of the Philosophy of Science Association (Vol. 1994, No. 2, pp. 88-98). Cambridge University Press.
\item Schwartz, P. K., \& Giulini, D. (2020). Classical perspectives on the Newton?Wigner position observable. International Journal of Geometric Methods in Modern Physics, 17(12), 2050176.
\item Schweber, S. S., \& Bethe, H. A. (1964). An introduction to relativistic quantum field theory (Vol. 136). New York: Harper \& Row.
\item Segal, I. (1964). Quantum fields and analysis in the solution manifolds of dif- ferential equations. In W. Martin and I. Segal (Eds.), Analysis in Function Space, Cambridge, Massachusets. MIT Press.
\item Sternberg, S. (1995). Group theory and physics. Cambridge University Press.
\item Streater, R. F. (1975). Outline of axiomatic relativistic quantum field theory. Reports on Progress in Physics, 38(7), 771.
\item Sudarshan, E. C. G., \& Mukunda, N. (1974). Classical dynamics: a modern perspective.
\item Wallace, D. (2001). Emergence of particles from bosonic quantum field theory. arXiv preprint quant-ph/0112149.
\item Weinberg, Steven. (1995). The quantum theory of fields. Vol. 2. Cambridge university press.
\item Wightman, A. S. (1962). On the localizability of quantum mechanical systems. Reviews of Modern Physics, 34(4), 845.
\item Wigner, E. P. (1931). Gruppentheorie und ihre Anwendung auf die Quantenmechanik der Atomspektren.
\item Wigner, E. (1939). On unitary representations of the inhomogeneous Lorentz group. Annals of mathematics, 149-204.
\item Wigner, E. P. (1963). The problem of measurement. American Journal of Physics, 31(1), 6-15.

\end{itemize}

\end{document}